\documentclass[11pt]{iopart} 
\usepackage{lipsum}
\usepackage{iopams}
\usepackage{amsmath,amsfonts,amssymb,mhchem}
\usepackage{graphicx}
\usepackage{siunitx}
\DeclareSIUnit{\molar}{M}
\usepackage{float}
\usepackage{lineno}
\usepackage{filecontents}

\usepackage{hyperref}
\hypersetup{hidelinks}
\usepackage{xr}
\usepackage[font=footnotesize,labelfont=bf]{caption}
\usepackage{wrapfig}
\usepackage[outercaption]{sidecap}
\makeatletter
\def\SC@figure@vpos{m}
\makeatother
\usepackage{tabularx}
\usepackage{color,soul}
\usepackage{bm}
\usepackage{titletoc}
\usepackage{caption}
\renewcommand{\figurename}{Fig.}
\usepackage{ulem}
\usepackage{matlab-prettifier}
\usepackage[backend=bibtex, style=nature]{biblatex}
\renewbibmacro{begentry}{\midsentence}
\AtBeginBibliography{\small}

\usepackage{etoolbox}
\usepackage{setspace}
\newcounter{suppfigure}
\newcommand{\listofsuppfigures}{\section*{Supplementary Figures}\@starttoc{lof}}

\DeclareCiteCommand{\citen}
  {}
  {\printfield{labelnumber}}
  {\multicitedelim}
  {}

\begin{document}
\singlespacing
\raggedbottom
\clearpage

\title[]{Electron--Ion Coupling Breaks Energy Symmetry in Bistable Organic Electrochemical Transistors}

\author{Lukas M. Bongartz$^{1,2*}$, Garrett LeCroy$^{1}$, Tyler J. Quill$^{1}$, Nicholas Siemons$^{1}$, Gerwin Dijk$^{1}$, Adam Marks$^{1}$, Christina Cheng$^{1}$, Hans Kleemann$^{2}$, Karl Leo$^{2}$, Alberto Salleo$^{1}$}

\address{$^{1}$Department of Materials Science and Engineering, Stanford University, Stanford, CA 94305, USA}
\address{$^{2}$Institute for Applied Physics, Technische Universit\"at Dresden, N\"othnitzer Str. 61, 01187 Dresden, Germany}

\begin{abstract}
Organic electrochemical transistors are extensively studied for applications ranging from bioelectronics to analog and neuromorphic computing. Despite significant advances, the fundamental interactions between the polymer semiconductor channel and the electrolyte, which critically determine the device performance, remain underexplored. Here, we examine the coupling between the benchmark semiconductor PEDOT:PSS and an ionic liquid to explain the bistable and non-volatile behavior observed in OECTs. Using X-ray scattering and spectroscopy techniques, we demonstrate how the electrolyte modifies the channel composition, enhances molecular order, and reshapes the energetic landscape. Notably, the observed bistability arises from asymmetric and path-dependent energetics during doping and dedoping, resulting in two distinct paths, driven by a direct interaction between the electronic and ionic charge carriers. These findings highlight the electrolyte's role in tuning charge carrier dynamics, positioning it as a powerful yet underutilized lever for enabling novel device functionalities.
\end{abstract}

\section*{Introduction}

\begin{figure}[t!]
    \centering
    \includegraphics[width=\linewidth]{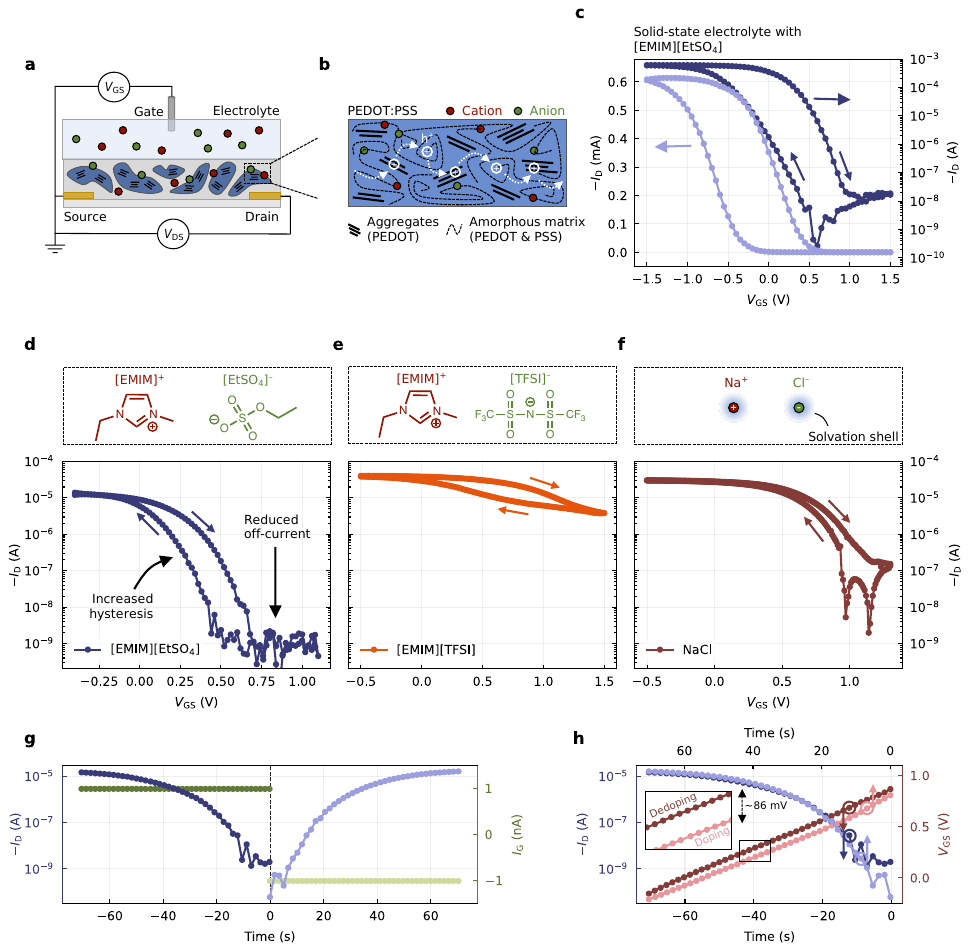} 
    \caption{\textbf{OECTs with PEDOT:PSS channel and \ce{[EMIM][EtSO4]} electrolyte.} (\textbf{a}) Generalized layout of an OECT with PEDOT:PSS channel. Devices in this work typically feature a capacitive side-gate (Fig.\,\ref{fig:S_Devices}). (\textbf{b}) PEDOT:PSS is a heterogeneous material, where the electronic charge transport through aggregated, paracrystalline entities is bridged by amorphous regions. (\textbf{c}) Transfer curve in linear and logarithmic scales of a bistable OECT based on PEDOT:PSS with an \ce{[EMIM][EtSO4]} solid-state electrolyte, as reported in ref.\,\citen{bongartz2024bistable} ($V_\mathrm{DS}=-200\,\si{\milli\volt}$). (\textbf{d}--\textbf{f}) Transfer curves of OECTs with PEDOT:PSS channel and plain \ce{[EMIM][EtSO4]}, \ce{[EMIM][TFSI]}, and $100\,\si{\milli\molar}$ aqueous \ce{NaCl} electrolyte ($V_\mathrm{DS}=-10\,\si{\milli\volt}$). (\textbf{g}) Operating an OECT (\ce{[EMIM][EtSO4]} electrolyte) with a gate current instead of a gate voltage ($V_\mathrm{DS}=-10\,\si{\milli\volt}$). (\textbf{h}) The folded $I_\mathrm{D}$-response shows no hysteresis. Instead, the hysteresis reflects in the potential difference between source and gate.
    \label{fig:1}}
     \vspace*{-\baselineskip}
\end{figure}

\noindent Organic electrochemical transistors (OECTs) are a core technology for next-generation electronic devices that couple ionic and electronic charge carriers\autocite{friedlein2018device, rivnay2018organic}. With functional similarities to biological neurons, OECTs hold promise for applications such as bioelectronics\autocite{cea2023integrated, rashid2021organic, nawaz2021organic, van2023retrainable}, scalable neuromorphic computing\autocite{fuller2019parallel, van2018organic, huang2023vertical}, and analog spiking neurons\autocite{sarkar2022organic, harikesh2023ion, zhu2023leaky}. Essential to OECTs is the channel material, where charge balance occurs inside organic mixed ionic--electronic conductors (OMIECs)\autocite{paulsen2020organic}, enabling ion density changes to modulate the electronic charge carrier concentration. This balance is effectively harnessed when an OMIEC acts as the channel material between source and drain contacts, while interfacing with the gate through an electrolyte (Fig.\,\ref{fig:1}a). A drain--source voltage $V_\mathrm{DS}$ drives electronic transport along the molecules' $\pi$-systems, while a gate--source voltage $V_\mathrm{GS}$ regulates ion transport between the electrolyte and channel. In maintaining charge neutrality, ions induce or compensate molecular charges, enabling dynamic de-/doping and control of the bulk conductivity.

Although a variety of systems have been explored as OECT channel materials\autocite{paulsen2020organic, tropp2023organic}, the most commonly deployed OMIEC is the polymer blend poly(3,4-ethylenedioxythiophene) polystyrene sulfonate (PEDOT:PSS, Fig.\,\ref{fig:S_PEDOTPSS}). Within this blend, PEDOT provides the $\pi$-backbone for electronic transport and is (p-)doped by the ionically conducting PSS. On the nano- and mesoscale, PEDOT:PSS forms cores, or ``pancakes''\autocite{kayser2019stretchable, ahmad2021mechanisms}, embedded in a PSS-rich matrix\autocite{takano2012pedot, rivnay2016structural, keene2022efficient}, coexisting in regions of paracrystalline order and amorphous disorder (Fig.\,\ref{fig:1}b). Extensive research has focused on fine-tuning the properties of PEDOT:PSS to suit various applications. For example, post-treatment with sulfuric acid (\ce{H2SO4}) improves the long-term stability by enhancing crystallinity and removing excess PSS\autocite{kim2018influence, kim2014highly}. PEDOT:PSS-based OECTs can be operated in enhancement mode by utilizing molecular dedoping to reduce off-current and threshold voltage\autocite{keene2020enhancement}. Alternatively, different additives have been explored to induce non-volatility in neuromorphic computing, manifesting as a markedly hysteretic switching performance\autocite{van2017non, ji2021mimicking}. Despite being effective, these approaches rely on additional processes and focus on modifying the semiconductor material. In that context, ionic liquids have gained increasing attention as water-free substitutes for the electrolyte, holding promise to improve the device performance in terms of temperature resilience and operation bandwidth\autocite{melianas2020temperature, zhong2023ionic}.

We have recently shown that PEDOT:PSS-based OECTs, when operated with an \ce{[EMIM][EtSO4]} electrolyte in a solid-state matrix, show a pronounced hysteretic switching performance (Fig.\,\ref{fig:1}c). This hysteresis is stable over consecutive cycles and persists even at strongly reduced scan rates\autocite{weissbach2022photopatternable, bongartz2024bistable}, hinting at a non-kinetic origin (Fig.\,\ref{fig:S_SolidState}). We previously described this phenomenon by a thermodynamic framework, in which such bistable device operation arises from the balance of entropic and enthalpic contributions involved in the doping cycle\autocite{bongartz2024bistable}. While this reasoning provides a description from a coarse-grained perspective, a microscopic understanding of the energetics in the system is still lacking. That is the objective of this paper. 

We begin by reviewing the device performance of the OECT system, and comparing it with two reference electrolytes: aqueous NaCl solution and the ionic liquid \ce{[EMIM][TFSI]}. We then perform a material study, revealing peculiar interactions between PEDOT:PSS and the \ce{[EMIM][EtSO4]} electrolyte, including substantial alterations in the semiconductor's composition and morphology. The final piece of our study concerns an in-operando analysis of the de-/doping operation by means of spectroelectrochemical measurements. By applying a vibronic fitting model, we extract insights about the energetics of the operation cycle, uncovering an energetic asymmetry between doping and dedoping. This behavior appears to stem from a distinct coupling between the electronic charge carriers of the OMIEC and the ionic, molecular charge carriers from the ionic liquid electrolyte. This study highlights the choice of electrolyte as a critical factor in tuning OECT performance, revealing an effective, yet underexplored approach to impart novel functionalities to electrochemical devices.

\section*{Results and Discussion}
\subsection*{Device Study}

The ionic liquid \ce{[EMIM][EtSO4]} alone, without solid-state matrix, imparts strongly bistable device performance when used as the electrolyte in PEDOT:PSS-based OECTs (Fig.\,\ref{fig:1}d). As in the solid-state system, this performance is persistent across cycles, retained at low scan rates, and goes along with low off- and gate currents (Fig.\,\ref{fig:S_Transfer_Cycling}). While also being present under ambient conditions, these features are considerably more stable when the device is operated in an inert-gas atmosphere, thus defining the standard measurement conditions herein. For reference, Fig.\,\ref{fig:1}e,f show the transfer curves of equivalent OECT devices, operated with the ionic liquid \ce{[EMIM][TFSI]} and the standard electrolyte of aqueous NaCl (100\,mM). Besides its pronounced hysteresis, the \ce{[EMIM][EtSO4]} device stands out for its low off-current in the range of $\si{\nano\ampere}$, which goes along with a reduction in threshold voltage. Hysteresis is similarly present with \ce{[EMIM][TFSI]}, yet at a markedly higher off-current level, and scarcely with \ce{NaCl}. Interestingly, when an OECT is temporarily exposed to \ce{[EMIM][EtSO4]} but operated in \ce{NaCl}, the changes to threshold voltage and off-current persist, reminiscent of sustained molecular dedoping (Fig.\,\ref{fig:S_NaCl_Exposure}). No hysteresis is observed with \ce{[EMIM][EtSO4]} when PEDOT:PSS is exchanged for another OMIEC (Fig.\,\ref{fig:S_Transfer}).

Evidently, a transfer curve hysteresis implies different channel conductivities during the doping and dedoping operation, for a given gate--source voltage. Such asymmetry may arise from both transport (mobility) and charge-carrier phenomena, which together determine the electronic conductivity. We measure the hole mobilities by a pulsed-current technique and find similar mobilities when the device is operated from doped and undoped initial states: $\mu_d = (12.2\pm0.6)\,\si{\square\centi\metre\per\volt\per\second}$ and $\mu_u = (11.6\pm0.6)\,\si{\square\centi\metre\per\volt\per\second}$ (\hyperref[Note_S:Mobility]{Supplementary Note 1}). This finding suggests that the hysteresis is mostly governed by the charge carrier density: in a transfer curve measurement, where the gate voltage controls the drain current, the charge carrier density, for a given gate--source voltage, depends on the doping history. Vice versa, it follows that for a given charge carrier density, the associated potential difference between source and gate is different, depending on whether the system approaches that state from a doped or undoped initial state. 

We validate this reasoning by means of an ``inverted'' transfer curve, where the number of charge carriers is controlled by a gate current $I_\mathrm{G}$, and the corresponding gate--source voltage is measured. Fig.\,\ref{fig:1}g shows the modulation in drain current upon applying $1\,\si{\nano\ampere}$ to the gate, followed by $-1\,\si{\nano\ampere}$ ($70\,\si{\second}$ each). Effective de-/doping is observed, with the drain current varying over four orders of magnitude and saturating in the on and off states. Fig.\,\ref{fig:1}h overlays the two sweeps: no hysteresis is present as the charge-carrier number is fixed. Instead, the hysteresis manifests in a constant offset of $85.6 \pm 0.5\,\si{\milli\volt}$ in the associated gate--source voltage $V_\mathrm{GS}$. This result suggests that dedoping is energetically less favorable than doping, apparently consistent with the ``clockwise'' hysteresis loop in Fig.\,\ref{fig:1}d. Charge carrier injection and retraction exhibit different energetics, suggesting changes in the energy landscape and density of states, which motivates an analysis of the semiconductor composition and morphology. 

\subsection*{Semiconductor Analysis}

Ionic liquids are known to induce a range of irregular effects in polymer semiconductors\autocite{li2024boosting, teo2017highly, wu2022ionic}. Kee et al. have identified a counter-ion exchange in the treatment of PEDOT:PSS with different ionic liquids, which goes along with a reorganization of PEDOT towards higher crystallinity\autocite{kee2016controlling}. We verify these effects for PEDOT:PSS channels that were temporarily exposed to \ce{[EMIM][EtSO4]} and \ce{[EMIM][TFSI]}.

\begin{figure}[t!]
    \centering
    \includegraphics[width=\linewidth]{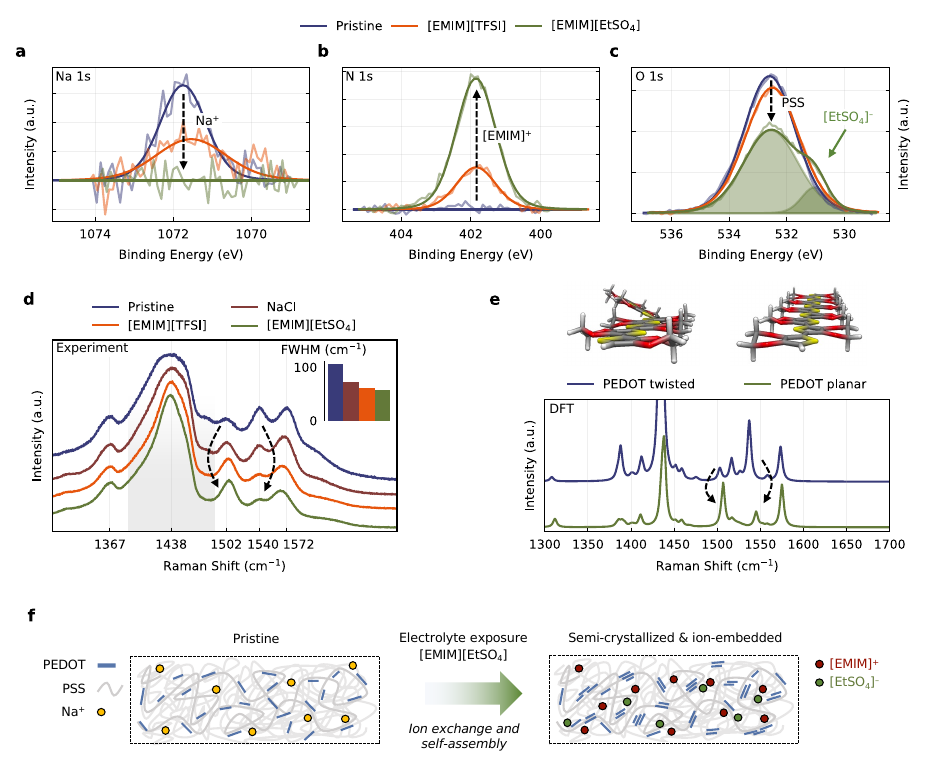} 
    \caption{\textbf{Semiconductor analysis.} (\textbf{a}--\textbf{c}) XPS spectra (raw and fitted) of pristine and electrolyte-treated PEDOT:PSS thin films with core levels (\textbf{a}) Na\,1s, (\textbf{b}) N\,1s, and (\textbf{c}) O\,1s. See Fig.\,\ref{fig:S_XPS_C_S} for C\,1s and S\,2p spectra. Raw data were fitted to Gauss--Lorentz profiles. (\textbf{d}) Raman spectra (high-energy range) measured with $\lambda_\mathrm{exc}=532\,\si{\nano\metre}$. The full spectra and peak assignment are provided in Fig.\,\ref{fig:S_Raman} and Tab.\,\ref{tab:S_Raman}. (\textbf{e}) DFT calculations of distorted and planar PEDOT reproduce the experimental observations in (\textbf{d}). (\textbf{f}) Schematic illustrating the effect of \ce{[EMIM][EtSO4]} on PEDOT:PSS. Along with an ion exchange, where PSS is removed and \ce{[EMIM]+} and \ce{[EtSO4]-} are retained, the PEDOT chains reorganize and self-assemble with higher order.
    \label{fig:2}}
     \vspace*{-\baselineskip}
\end{figure}

XPS analysis reveals a reduction in the Na\,1s signal by both electrolytes (Fig.\,\ref{fig:2}a), which is accompanied by a rise in the N\,1s signal (Fig.\,\ref{fig:2}b). The C\,1s data (Fig.\,\ref{fig:S_XPS_C_S}), lacking a \ce{CF3} signal owing to \ce{[TFSI]-}, implies that the nitrogen signal is attributed to \ce{[EMIM]+} alone. These results confirm a cation exchange, where residual, PSS-bound sodium ions\autocite{cho2018influence, lyu2023operando} are exchanged by the molecular \ce{[EMIM]+} cations from the electrolytes. Noteworthy, this exchange appears complete for the treatment with \ce{[EMIM][EtSO4]}, while being only partial with \ce{[EMIM][TFSI]}. The O\,1s core-level region (Fig.\,\ref{fig:2}c) further highlights a PSS removal, which is much more pronounced with \ce{[EMIM][EtSO4]}, and where it goes along with a distinctive shoulder at $\sim531\,\si{\electronvolt}$. This shoulder is attributed to the \ce{[EtSO4]-} anion\autocite{gottfried2006surface, jurado2016effect}, suggesting that \ce{[EMIM]+} is not only bound to PSS, but is also partially balanced by its native anion \ce{[EtSO4]-}. With \ce{[EMIM][TFSI]}, the cation is retained independently of the anion. 

We attribute this interplay between PEDOT:PSS and \ce{[EMIM][EtSO4]} to two main factors. First, \ce{[EMIM][EtSO4]} stands out by a markedly high dielectric constant ($\epsilon_{\ce{[EMIM][EtSO4]}}=27.9$ vs. $\epsilon_{\ce{[EMIM][TFSI]}}=12.3$)\autocite{weingartner2006static, singh2008static}, enabling it to penetrate the thin-film and shield the interactions between PEDOT and PSS, allowing the polymers to separate\autocite{de2018ionic, de2021ionic}. Second, the chemical similarity between \ce{[EtSO4]-} and PSS, both containing \ce{SO3}-groups, suggests comparable binding energetics towards PEDOT and \ce{[EMIM]+}. In fact, DFT studies by de Izarra et al. estimate the free energy of ion exchange between PEDOT:PSS and \ce{[EMIM][EtSO4]} as $<20\,\si{\milli\electronvolt}$\autocite{de2018ionic}. Assuming a Boltzmann distribution, this suggests an ion-exchange ratio of $\sim32\%$ at room temperature. We estimate this figure from the fitted XPS data and find a comparable ratio of 24\% for PEDOT:PSS with \ce{[EMIM][EtSO4]} (\hyperref[Note_S:Ion_Transfer]{Supplementary Notes 2 and 3}).

GIWAXS analysis is performed on thin films of equivalent treatment, reflecting the removal of PSS for both films exposed to an ionic liquid (Fig.\,\ref{fig:S_GIWAXS}). An increasing PEDOT $\pi$-stacking peak indicates slightly enhanced crystallinity with \ce{[EMIM][EtSO4]}, aligning with the results of Kee et al.\autocite{kee2016controlling}. Meanwhile, thin-film UV-Vis and IR spectroscopy reveal no significant changes across the systems, except for the sample treated with \ce{[EMIM][EtSO4]}, which exhibits a more pronounced polaron absorbance and a vibrational signature attributed to the retained \ce{[EMIM]+} cation (Fig.\,\ref{fig:S_UVVis}, \ref{fig:S_IR}).

As a complementary technique, we use Raman spectroscopy to probe modes that are weak or inactive in IR (Fig.\,\ref{fig:2}d, \ref{fig:S_Raman}). The principal peak of PEDOT ($1438\,\si{\per\centi\metre}$) undergoes no significant shift with any electrolyte exposure ($<4\,\si{\per\centi\metre}$). However, we identify a meaningful decrease in bandwidth, with \ce{[EMIM][EtSO4]} reducing the full width at half maximum (FWHM) from $106\,\si{\per\centi\metre}$ to $57\,\si{\per\centi\metre}$ (Fig.\,\ref{fig:S_Raman_FWHM}). Given that the peak bandwidth reflects the statistical distribution of conjugation lengths\autocite{kong2022advanced}, we interpret this as evidence of \ce{[EMIM][EtSO4]} arranging and homogenizing PEDOT. The higher-sided signals in Fig.\,\ref{fig:2}d, assigned to asymmetric mid- and end-chain modes, meanwhile exhibit a peculiar intensity shift. The $1572\,\si{\per\centi\metre}$ signal remains largely unchanged, while the $1502\,\si{\per\centi\metre}$ mode increases and the $1540\,\si{\per\centi\metre}$ mode nearly vanishes with \ce{[EMIM][EtSO4]}. Similar patterns have been reported for the treatment with sulfuric acid\autocite{jucius2019structure}, which affirms the homogenizing effect on PEDOT:PSS. To support this conclusion, we perform DFT calculations of the Raman signals, with PEDOT in distorted and planar configurations (\hyperref[Note_S:Raman_DFT]{Supplementary Note 4}). Shown in Fig.\,\ref{fig:2}e, the calculations agree with the experimental data, demonstrating that the transition from a distorted to a planar configuration manifests in the relative intensity shift. 

Fig.\,\ref{fig:2}f summarizes the effects of \ce{[EMIM][EtSO4]} on PEDOT:PSS. Upon electrolyte exposure, the system undergoes a thermally activated ion exchange, where residual sodium ions are replaced by \ce{[EMIM]+}, and both \ce{[EMIM]+} and \ce{[EtSO4]-} are retained in the semiconductor film. Meanwhile, a substantial share of PSS is removed, allowing the PEDOT chains to reorganize and self-assemble into higher-order structures. We infer that this removal of PSS drives the exceptional reduction in OECT off-currents (Fig.\,\ref{fig:1}c,d), which can be motivated by two mechanisms: First, interlocked \ce{PSS} can act as a parasitic dopant, contributing to off-state conductivity in terms of charge carrier density. Second, \ce{PSS} supports the bulk conductivity in PEDOT:PSS by facilitating electronic tunneling\autocite{keene2022efficient}, contributing to off-state conductivity in terms of charge carrier transport. Finally, the high polarity of \ce{[EMIM][EtSO4]} facilitates film penetration and polymer access for dedoping, which does not hold for the less polar \ce{[EMIM][TFSI]}.

Particularly striking is the enduring presence of \ce{[EtSO4]-} anions in the film, where they complement \ce{PSS} in counterbalancing and stabilizing positively charged holes (polarons) on PEDOT. We hypothesize that this dual‐ion stabilization relates to the bistable switching performance in OECTs, which we assess by studying the doping energetics spectroscopically.

\subsection*{In-Operando Doping Studies}

\begin{figure}[t!]
    \centering
    \includegraphics[width=\linewidth]{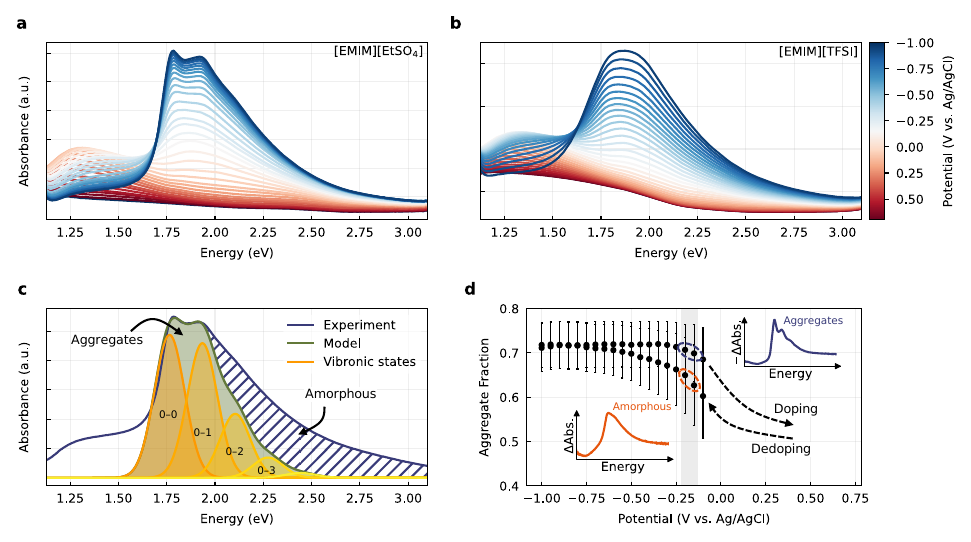} 
    \caption{\textbf{In-operando doping studies.} (\textbf{a}, \textbf{b}) Absorbance spectra (discharging cycle) of PEDOT:PSS with \ce{[EMIM][EtSO4]} and \ce{[EMIM][TFSI]} electrolyte. \ce{[EMIM][EtSO4]} produces feature-rich spectra, reflecting high aggregation, while \ce{[EMIM][TFSI]} suggests an amorphous nature. (\textbf{c}) Absorbance spectra are fitted using a vibronic model that incorporates the lowest five vibronic transitions (0--0 through 0--4). This approach allows to infer the absorbance share owing to undoped PEDOT in aggregates and amorphous regions. (\textbf{d}) From this, the aggregate fraction ($AF$) follows for each potential. The omitted potential regime cannot be reliably fitted due to dominant polaron absorbance (\hyperref[Note_S:Spectroelectrochemistry_Fits]{Supplementary Note 5}). Insets: Differential absorbance for the dis-/charging step between potentials of $-0.2\,\si{\volt}$ and $-0.15\,\si{\volt}$, revealing a stark asymmetry in the spectral signatures. \label{fig:3}}
     \vspace*{-\baselineskip}
\end{figure}

Charging the polymer semiconductor to different doping degrees via chronoamperometry (Fig.\,\ref{fig:S_CA}), while simultaneously measuring its spectral features, provides insights into the morphological environments and energetics involved in de-/doping. Fig.\,\ref{fig:3}a and b show the discharging cycle of PEDOT:PSS with \ce{[EMIM][EtSO4]} and \ce{[EMIM][TFSI]} electrolyte (see Fig.\,\ref{fig:S_Spec_EMIMEtSO4} and \ref{fig:S_Spec_EMIMTFSI} for both cycles). The ionic liquid \ce{[EMIM][EtSO4]} induces a feature-rich absorbance signature of PEDOT, resembling the \ce{NaCl} sample (Fig.\,\ref{fig:S_Spec_NaCl}) more closely than \ce{[EMIM][TFSI]}. The dedoped states below $1.6\,\si{\electronvolt}$ show strong suppression of polaronic excitation, which directly correlates with the reduced off-current in OECTs. Absorbance above $1.6\,\si{\electronvolt}$ originates from excitonic transitions involving states of coupled electronic and vibrational coordinates (Fig.\,\ref{fig:S_Transition}), where amorphous regions broaden the density of states, while aggregates narrow it\autocite{kasha1963energy, hestand2018expanded}. Accordingly, the sample with \ce{[EMIM][TFSI]} indicates a considerably more amorphous morphology of PEDOT. The increased intensity of the 0--0 transition relative to the 0--1 transition for the \ce{[EMIM][EtSO4]} sample further implies dominant intrachain coupling. This contrasts with the \ce{NaCl} sample (Fig.\,\ref{fig:S_Spec_NaCl}), where more balanced intensities suggest a mix of intra- and interchain coupling\autocite{spano2005modeling, hestand2018expanded}. This observation resonates with the crystallizing impact of \ce{[EMIM][EtSO4]} on PEDOT found earlier. 

 \begin{figure}[t!]
    \centering
    \includegraphics[width=\linewidth]{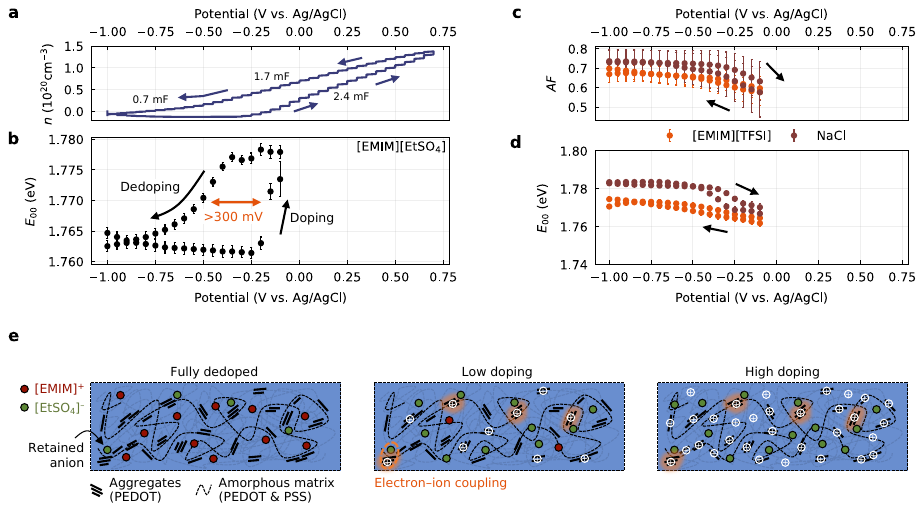} 
    \caption{\textbf{Energetics of the doping cycle.} (\textbf{a}) Charge carrier density ($n$) as a function of applied potential, calculated by integrating the current with respect to time (Fig.\,\ref{fig:S_CA}). See also Fig.\,\ref{fig:S_CA_Diff_EMIMEtSO4}. (\textbf{b}) $E_{00}$ energies extracted from absorbance fits (Fig.\,\ref{fig:3}c), showing a marked asymmetry between doping and dedoping. (\textbf{c}, \textbf{d}) Aggregate fraction ($AF$) and $E_{00}$ energies for samples with \ce{[EMIM][TFSI]} and \ce{NaCl} electrolyte. (\textbf{e}) Schematic illustration of the ionic and electronic population during doping. Left: Fully dedoped state; abundant \ce{[EMIM]+} and only residual \ce{[EtSO4]-} in ordered domains. Center: low-doping regime; initial hole injection into highly ordered PEDOT aggregates, partially co-stabilized by \ce{[EtSO4]-}. This stabilization must be overcome during dedoping. Right: high-doping regime; after aggregate saturation, holes extend into the amorphous phase.
    \label{fig:4}}
     \vspace*{-\baselineskip}
\end{figure} 

The absorbance of polymer semiconductors can be described by modified Holstein Hamiltonians, and in particular the model by Spano, which captures the vibronic features of crystalline aggregates\autocite{spano2005modeling, lecroy2023role}. We apply this formalism to the spectral data obtained with \ce{[EMIM][EtSO4]} (\hyperref[Note_S:Spectroelectrochemistry_Fits]{Supplementary Note 5}), isolating the absorbance stemming from PEDOT aggregates (Fig.\,\ref{fig:3}c). The remaining signal above $\sim1.6\,\si{\electronvolt}$ arises from amorphous domains. By integrating these two contributions, we extract an aggregate fraction ($AF$)---the ratio of undoped PEDOT in aggregates to the total undoped PEDOT---at each applied potential. We restrict our analysis to the low‐charge-carrier regime, here taken for potentials below $-0.1\,\si{\volt}$, above which the spectra become dominated by polaronic absorbance, preventing a reliable application of the model. Still, tracing $AF$ over the de-/doping cycle shows a clear hysteresis (Fig.\,\ref{fig:3}d). Its decrease during doping indicates that charges first populate states in aggregated environments, consistent with the oxidation potential being closer to midgap than that of amorphous species\autocite{kaake2010intrinsic, keene2023hole}. During dedoping, these same aggregates release their charge last and crucially, only at a more negative potential than required for their initial charging, giving rise to the observed hysteresis.

These energetics also reflect in the differential spectra (Fig.\,\ref{fig:3}d, insets). Representing the absorbance lost during doping or gained during dedoping, differentials indicate the environment where charges are injected or extracted. At the start of doping, feature-rich spectra show prioritized charging of aggregates, which gradually blur out as the potential rises and amorphous regions follow. This sequence is reversed for dedoping, but occurs with a distinct offset in potential. The insets of Fig.\,\ref{fig:3}d both correspond to the potential step from $-0.2\,\si{\volt}$ to $-0.15\,\si{\volt}$, highlighting that, at this potential, doping predominantly charges ordered regions, whereas dedoping mainly affects amorphous sites. Discharge of aggregates begins only at potentials below $-0.35\,\si{\volt}$, which links to the findings from Fig.\,\ref{fig:1}h: a higher potential is required to extract charges than to inject them.
\\ \\ \noindent This energetic asymmetry also reflects in the charge carrier density ($n$) in Fig.\,\ref{fig:4}a (Fig.\,\ref{fig:S_CA_Diff_EMIMEtSO4}). Charge injection begins at around $-0.3\,\si{\volt}$, with a steeper slope than charge extraction in the same regime ($2.4\,\si{\milli\farad}$ vs. $1.7\,\si{\milli\farad}$). Interestingly, the discharging curve further kinks at around $-0.5\,\si{\volt}$ ($0.7\,\si{\milli\farad}$), which suggests a bimodal stabilization of charge carriers.

From the model fits, we also evaluate the evolution of the $E_{00}$ parameter, shown in Fig.\,\ref{fig:4}b. Representing the center of the 0--0 Gaussian (Fig.\,\ref{fig:3}c), it corresponds to the transition energy between the fundamental phonon modes of the ground and excited states in undoped PEDOT. Upon the onset of doping, $E_{00}$ increases gradually. This rise is reversed under dedoping, but only with a potential shift of more than $300\,\si{\milli\volt}$. Notably, in this regime, the curve also reflects the bimodal stabilization through a flattening slope at around $-0.6\,\si{\volt}$. We interpret this result as evidence for multiple ordered species participating in the doping cycle. More ordered domains are expected to exhibit lower $E_{00}$ values than disordered ones, owing to greater coherence. Thus, the evolution of $E_{00}$ throughout the cycle aligns with our aggregate‐fraction data: charges first occupy highly ordered aggregates, then fill less ordered, and finally amorphous regions.

For comparison, Fig.\,\ref{fig:4}c,d present the aggregate fraction and $E_{00}$ evolution for samples using \ce{[EMIM][TFSI]} and \ce{NaCl} as electrolytes. While the aggregate fraction follows a similar trend, the extent of hysteresis is notably smaller. Meanwhile, none of the previously described effects appear in the $E_{00}$ evolution; we observe only a minimal decrease during doping and, crucially, no comparable shift along the potential axis under dedoping. This behavior resonates with our XPS study, which identified \ce{[EMIM][EtSO4]} as uniquely embedding its anion due to favorable binding energetics with PEDOT:PSS, suggesting its presence as a key driver behind the energetic asymmetry. 
\\ \\ \noindent Specifically, we understand \ce{[EtSO4]-} to stabilize polarons within highly ordered PEDOT aggregates, which are the first sites to accept charges during doping (Fig.\,\ref{fig:4}e). Its small ionic radius enables \ce{[EtSO4]-} to access cavities that are inaccessible to PSS, thereby lowering the energy of these doped species and introducing an energetic bias between the forward and reverse operation. This mechanism also aligns with the strong dependence of the OECT hysteresis on the gate voltage scan range (Fig.\,\ref{fig:S_ScanRange})\autocite{weissbach2022photopatternable}. For shallow dedoping, charge exchange remains confined to amorphous regions and no hysteresis is observed. For deep dedoping, it extends into the ordered aggregate phase, where \ce{[EtSO4]-} impedes charge extraction, driving the development of hysteresis. 

We interpret this mechanism as a manifestation of the distinctive alignment between the electronic charge carriers of the polymer semiconductor and the ionic charge carriers of the ionic liquid electrolyte, specifically with respect to dipole moment and polarizability: crucial properties that drive charge-carrier coupling beyond simple Coulombic interaction. In both doped polymer semiconductors and molecular ions of ionic liquids, electronic states are delocalized over a molecular $\pi$-system, leading to comparable effects on dipole moment and polarizability. This contrasts sharply with atomic ions such as chloride, which possess neither a permanent dipole moment nor significant polarizability, and are further enclosed within a solvation shell, impeding coupling with the doped OMIEC. Accordingly, less stabilization of the doped state is expected than with molecular ions such as \ce{[EtSO4]-}.

We hypothesize that this process may also involve subtle molecular rearrangements of \ce{[EtSO4]-}, potentially orienting toward charged PEDOT aggregates. Tentative evidence for such rearrangements may be found in the differential absorbance spectra, which exhibit a clear asymmetry in the 0--0/0--1 transition ratio between doping and dedoping, indicative of a disruption in the local electronic environment (Fig.\,\ref{fig:S_CA_Diff_EMIMEtSO4}). However, further verification of this hypothesis is challenging, as in-operando imaging techniques remain less accessible for PEDOT:PSS as compared to systems with stronger diffraction\autocite{quill2023ordered}.

\section*{Conclusion}

We studied the performance of OECTs using the benchmark semiconductor PEDOT:PSS with the ionic liquid electrolyte \ce{[EMIM][EtSO4]}. This combination has previously been shown to exhibit peculiar device characteristics, including a significantly suppressed off-current and pronounced non-volatile, hysteretic behavior\autocite{bongartz2024bistable}. We find that \ce{[EMIM][EtSO4]} induces multiple effects in PEDOT:PSS, including the removal of excess PSS dopant material and the retention of the \ce{[EtSO4]-} anion. Its electronic properties and small size enable \ce{[EtSO4]-} to stabilize polarons in its vicinity, which creates an energetic asymmetry between doping and dedoping. While doping is energetically favored, dedoping requires overcoming this stabilization, resulting in distinct switching energetics.

These results underscore the critical interplay between the components of electrochemical devices, particularly the direct interaction between the electronic charge carriers in the OMIEC and the ionic charge carriers supplied by the electrolyte. This coupling is especially pronounced in polar ionic liquids, whose molecular ions match the electronic structure of the charged OMIEC in terms of dipole moment and polarizability, promoting a direct coupling of their electron densities. Conventional electrolytes like aqueous NaCl hinder such coupling due to their electronically isotropic, atomic ions and interfering solvation shells, a contrast often masked by the blanket designation of both as ``ionic charge carriers.'' While substantial effort has been directed toward optimizing OMIECs, the electrolyte design space, particularly for molecular systems, remains largely untapped. As this study shows, it holds great promise to impose novel functionalities to electrochemical systems. Further understanding of the precise coupling mechanisms between doped polymer semiconductors and charged molecular ions will unlock new degrees of freedom, enabling functionalities beyond the reach of conventional electronic materials, with implications extending beyond the specific system examined here. Unlocking such dynamics offers opportunities to, among others, advance post-binary computing, enabling systems founded on complex behaviors and multi-state stability\autocite{parisi2023nobel}.

\section*{Methods}

\noindent \textbf{Device fabrication.} Except for the solid-state devices (see below), microfabrication of OECTs was performed on silicon wafer substrates with a $1\,\si{\micro\metre}$ thermal oxide, employing previous established protocols\autocite{dijk2023pedot} that involve a lift-off process to pattern the source and drain contacts and an additional dry lift-off with a sacrificial Parylene C layer to pattern PEDOT:PSS. The metal interconnects were patterned with photolithography using an LOL2000/SPR3612 bilayer resist and a maskless aligner (Heidelberg MLA150) followed by the deposition of $5/100\,\si{\nano\metre}$ Ti/Au (AJA e-beam evaporator) and lift-off. Patterning of the Parylene C encapsulation and PEDOT:PSS (channel and gate) were performed in a combined process through a dry lift-off process using a sacrificial Parylene C layer. First, $2\,\si{\micro\metre}$ Parylene C was deposited using 3‐(trimethoxysilyl)propyl methacrylate as an adhesion promotor (Special Coating Systems PDS 2010 Labcoter). After the deposition of a diluted Micro-90 anti-adhesion layer, a second layer of $2\,\si{\micro\metre}$ Parylene C was deposited. Channel and gate were defined by etching the two layers of Parylene C (Plasma-Therm oxide etcher) using a $75\,\si{\nano\metre}$ Ti etch mask that was patterned using SPR 3612 photoresist and a metal etcher (Plasma-Therm metal etcher). PEDOT:PSS was deposited by preparing a solution of Hereaus Clevios PH1000, 6\,vol\% ethylene glycol (EG), 1\,vol\% (3-glycidyloxypropyl)trimethoxysilane (GOPS), and 0.1\,vol\% dodecyl benzene sulfonic acid (DBSA), all of which were obtained from Sigma Aldrich. The solution was spun at 1000\,rpm for $2\,\si{\minute}$ and baked at $120\,\si{\celsius}$ for $5\,\si{\minute}$. The sacrificial Parylene C layer was peeled off, followed by an additional baking for $15\,\si{\minute}$. Dies were rinsed in deionized water and dried for another $5\,\si{\minute}$ at $120\,\si{\celsius}$. The devices used in this work had dimensions of $W = 100\,\si{\micro\metre}, L = 100\,\si{\micro\metre}$, and a channel thickness of $100\,\si{\nano\metre}$. The gate area was $3.8 \times 2.2\,\si{\square\milli\metre}$. 

Solid-state devices were fabricated by the methods laid out in ref.\,\citen{bongartz2024bistable}. OECTs were fabricated on $1\,\text{inch} \times 1\,\text{inch}$ glass substrates with Cr/Au ($3\,\mathrm{nm}/50\,\mathrm{nm}$) traces deposited by evaporation. AZ 1518 photoresist was spin-coated ($3000\,\mathrm{rpm}$ for $60\,\si{\second}$), baked ($110\,\si{\celsius}$ for $60\,\si{\second}$), patterned by photolithography (I-line $365\,\mathrm{nm}, 167\,\mathrm{W}$ for $10\,\si{\second}$), developed (AZ 726 MIF), and Au/Cr were etched ($60\,\si{\second} / 20\,\si{\second}$) in 10\% aqueous etchants. After \ce{O2}-plasma cleaning, PEDOT:PSS (PH1000, with 5\% v/v ethylene glycol) was spin-coated ($3000\,\mathrm{rpm}$ for $60\,\si{\second}$, $\sim 100\,\mathrm{nm}$), dried ($120\,\si{\celsius}$ for $20\,\si{\minute}$), and patterned using OSCoR 5001 photoresist, which was spin-coated ($3000\,\mathrm{rpm}$ for $60\,\si{\second}$), baked ($100\,\si{\celsius}$ for $60\,\si{\second}$), exposed ($12\,\si{\second}$) to define channel and gate, post-baked ($100\,\si{\celsius}$ for $60\,\si{\second}$), and developed twice with Orthogonal Developer 103a. Excess PEDOT:PSS was removed by \ce{O2}-plasma etching ($5\,\si{\minute}, 0.3\,\mathrm{mbar}$). The resist was stripped overnight (Orthogonal Stripper 900), followed by ultrasonic ethanol cleaning ($15\,\si{\minute}$). For the solid-state electrolyte, the precursor solution was prepared by mixing deionized water ($1.0\,\mathrm{mL}$), N-isopropylacrylamide ($750\,\si{\milli\gram}$), N,N'-methylenebisacrylamide ($20\,\si{\milli\gram}$), 2-hydroxy-4'-(2-hydroxyethoxy)-2-methylpropiophenone ($200\,\si{\milli\gram}$), and 1-ethyl-3-methylimidazolium ethyl sulfate ($1.5\,\mathrm{mL}$), followed by overnight stirring at room temperature. Samples were treated with 5\% v/v 3-(trimethoxysilyl)propyl methacrylate in buffered ethanol (10\% v/v acetic acid/acetate) at $50\,\si{\celsius}$ for $10\,\si{\minute}$, rinsed in ethanol, and dried at $100\,\si{\celsius}$ for $15\,\si{\minute}$. The precursor was drop-cast, covered with Teflon\textsuperscript{TM} foil, and patterned in a mask aligner photolithography system ($20\,\si{\second}$ exposure). Uncrosslinked material was finally removed with \ce{N2}. Solid-state devices featured dimensions of $L = 30\,\si{\micro\metre}$ and $W = 150\,\si{\micro\metre}$.

\noindent \textbf{Electrical characterization.} Electrical characterizations were performed in an inert-gas atmosphere, using a Keithley 2612 SourceMeter with custom LabVIEW and SweepMe! software.

\noindent \textbf{XPS measurements.} 
PEDOT:PSS thin films of the same size were exposed to the ionic liquids \ce{[EMIM][EtSO4]} and \ce{[EMIM][TFSI]} for $5\,\si{\minute}$, and the excess electrolyte was removed by rising with deionized water. X-ray photoelectron spectroscopy (XPS) was performed using a PHI VersaProbe 4 with a monochromatized Al K$\alpha$ source ($1486.6\,\si{\electronvolt}$, $50\,\si{\watt}$, $200\,\si{\micro\metre}$ spot size) and a pass energy of $55\,\si{\electronvolt}$. An electron flood gun and low-energy \ce{Ar+} ions were used to neutralize the sample and prevent substrate charging. Binding energies were corrected to the \ce{C-C} peak in the C\,1s signal at $284.8\,\si{\electronvolt}$. For all collections, the angle between the sample surface and the detector was $45\si{\degree}$. To ensure that the compositions of the measured spectra were representative of the bulk, samples were sputtered with a gas cluster ion beam source (Ar$_{2000}\,^+$, $10\,\si{\kilo\volt}$) for $1\,\si{\minute}$ prior to all collections. The spectra were smoothed using the five-point quadratic Savitzky-Golay method and then fit to Gaussian-Lorentzian lineshapes and a Shirley-type background in CasaXPS software. 

\noindent \textbf{GIWAXS measurements.} Grazing incidence wide-angle X-ray scattering (GIWAXS) measurements were carried out at the Stanford Synchrotron Radiation Lightsource (SSRL) on beam line 11-3 using a CCD area detector (Rayonix MAR 225) at a distance of $319.998\,\si{\milli\metre}$ and an incidence angle of $0.1\si{\degree}$. The incident energy was $12.73\,\si{\kilo\electronvolt}$. The beam path was flooded with helium to prevent air scattering. Data were corrected for geometric distortion arising from the flat detector architecture and normalized by exposure time and monitor counts. Analysis was performed using Nika 1D SAXS\autocite{ilavsky2012nika} and WAXS tools\autocite{oosterhout2017mixing} software in Igor Pro. The reported quantities $q_{xy}$ and $q_{z}$ refer to the in-plane and out-of-plane scattering vector relative to the substrate. $q_{z}$ lineouts were taken from $\chi$-slices from $-20\si{\degree}$ to $20\si{\degree}$ with respect to the $q_{z}$ direction. Samples featured similar sizes, allowing for a semi-quantitative comparison of intensities.

\noindent \textbf{DFT calculations.} Density functional theory (DFT) calculations for the Raman experiment were carried out using the Gaussian16 software package\autocite{g16} with a B3LYP\autocite{becke1988density, lee1988development, slater1974self} functional and 6-311G basis set. The incident laser energy was matched to the experiment ($2.329\,\si{\electronvolt}$). Details are provided in \hyperref[Note_S:Raman_DFT]{Supplementary Note 4}.

\noindent \textbf{UV-Vis-IR spectroscopy.} Solid-state UV-Vis-IR spectroscopy was performed using an Agilent Cary 6000i UV-Vis-NIR Spectrophotometer and a Nicolet iS50R Fourier Transform Infrared (FTIR) spectrometer. For sample preparation, PEDOT:PSS was spun onto \ce{CaF2} substrates (2000\,rpm for $2\,\si{\minute}$) and immersed in the electrolyte for $5\,\si{\minute}$, followed by rinsing in deionized water. 

\noindent \textbf{Raman spectroscopy.} Raman spectroscopy was performed using a Horiba Labram HR Evolution confocal Raman microscope with an Andor CCD detector. A continuous wave $532\,\si{\nano\metre}$ excitation laser was used and Raman scatter was collected through an objective in backscattering geometry (100$\times$, 0.6 NA). The scattered light was passed through a long-pass filter ($\sim100$ cm$^{-1}$) and dispersed with a $1800\,\mathrm{gr}\,\si{\per\milli\metre}$ grating.

\noindent \textbf{Spectroelectrochemistry.} Spectroelectrochemical measurements were carried out using a home-built spectrometer with an Ocean Optics light source (tungsten-halogen light source) and an Ocean Optics QEPro detector. A quartz cuvette ($20\,\si{\milli\metre}$ optical path length) was used to conduct measurements in transmission mode. For sample preparation, PEDOT:PSS was spun onto indium-tin-oxide-coated glass substrates (ITO), covering $2.7\,\si{\square\centi\metre}$ in a thickness of $380\,\si{\nano\metre}$, as verified by profilometry (Bruker Dektak XT profilometer). Samples were placed in the cuvette filled with electrolyte to form a three-electrode cell with a Pt mesh counter electrode, an eDAQ leakless Ag/AgCl reference electrode, and the PEDOT:PSS-coated ITO substrate working electrode. The cuvette was capped and purged with Ar gas. The electrochemical potential was controlled using an Ivium CompactStat potentiostat (chronoamperometry). The potential at the working electrode was modulated in increments of $50\,\si{\milli\volt}$ between $-1.0\,\si{\volt}$ and $0.7\,\si{\volt}$, holding each potential for $30\,\si{\second}$. Spectra were recorded every $100\,\si{\milli\second}$ with an integration time of $10\,\si{\milli\second}$, and averaged over $5\,\si{\second}$ at the end of each potential step to best represent the steady state. Details on the vibronic fits are provided in \hyperref[Note_S:Spectroelectrochemistry_Fits]{Supplementary Note 5}.

\section*{Data Availability}

The data that support the findings of this study are available from the corresponding author upon reasonable request.

\section*{Code Availability}

The code underlying the vibronic fits is publicly available at \url{https://github.com/lukasbongartz/vibronic-fitting-tool.git}.

\section*{Acknowledgments}
L.M.B. acknowledges the German Academic Exchange Service (DAAD) and the Graduate Academy TU Dresden for support. 

\section*{References}
\printbibliography[heading = none]

\section*{Author Contributions}
L.M.B. and A.S. conceived the study. L.M.B. performed the electrical characterizations. T.J.Q. performed the XPS characterization. L.M.B. and G.L. carried out any optical spectroscopy. N.S. performed the DFT simulations. A.M. and C.C. carried out the GIWAXS measurement. G.D. fabricated the OECT devices. L.M.B. carried out the vibronic fitting and general data analysis. H.K. and K.L. provided guidance on data interpretation. L.M.B. wrote the manuscript, with input from G.L., T.J.Q., and G.D. All authors contributed to manuscript editing. 

\section*{Additional Information}

Additional information is given in the Supplementary Information. 

\section*{Competing Financial Interests}

The authors declare no competing interests.

\end{document}


\singlespacing
\raggedbottom
\clearpage

\title[]{Electron--Ion Coupling Breaks Energy Symmetry in Bistable Organic Electrochemical Transistors}

\small{\centering
\author{Lukas M. Bongartz$^{1,2*}$, Garrett LeCroy$^{1}$, Tyler J. Quill$^{1}$, Nicholas Siemons$^{1}$, Gerwin Dijk$^{1}$, Adam Marks$^{1}$, Christina Cheng$^{1}$, Hans Kleemann$^{2}$, Karl Leo$^{2}$, Alberto Salleo$^{1}$}

\address{$^{1}$Department of Materials Science and Engineering, Stanford University, Stanford, CA 94305, USA}
\address{$^{2}$Institute for Applied Physics, Technische Universit\"at Dresden, N\"othnitzer Str. 61, 01187 Dresden, Germany}}

\section*{\fontsize{14pt}{16pt}\selectfont \centering Supplementary Information}

\titlecontents{section}[0em]{}{\bfseries}{\bfseries}{}[\medskip]
\titlecontents{subsection}[1em]{}{}{}{\dotfill}[]

\startcontents[section]
\printcontents[section]{}{1}{\section*{Contents}}

\setcounter{equation}{0}
\renewcommand{\theequation}{S\arabic{equation}}
\setcounter{figure}{0}
\renewcommand{\thefigure}{S\arabic{figure}}
\setcounter{table}{0}
\renewcommand{\thetable}{S\arabic{table}}

\clearpage

\section*{Supplementary Figures}
\addcontentsline{toc}{section}{Supplementary Figures}

\begin{figure}[H]
    \centering
    \includegraphics[width=0.4\linewidth]{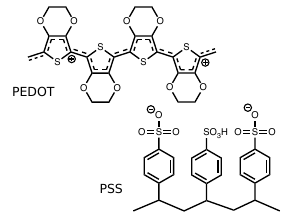} 
    \caption{\textbf{Molecular structure of PEDOT:PSS.} PEDOT:PSS is a blend of the electronic conductor poly(3,4-ethylenedioxythiophene) (PEDOT) doped with the polyanion and ionic conductor polystyrene sulfonat (PSS). \label{fig:S_PEDOTPSS}}
     \vspace*{-\baselineskip}
\end{figure}

\begin{figure}[H]
    \centering
    \includegraphics[width=\linewidth]{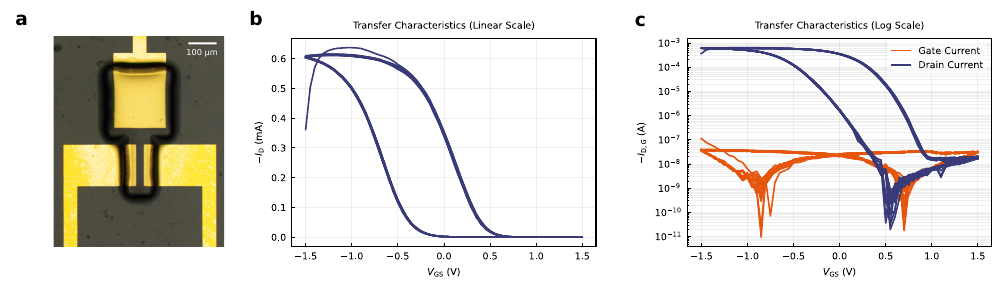}\caption{\textbf{Transfer characteristics for a bistable, solid-state OECT.} (\textbf{a}) Micrograph. The OECT uses a solid-state electrolyte based on \ce{[EMIM][EtSO4]}, as described in refs.\,\citen{weissbach2022photopatternable, bongartz2024bistable}. Transfer measurements in (\textbf{b}) linear scale and (\textbf{c}) logarithmic scale. Device dimensions: $W = 150\,\si{\micro\metre}, L = 30\,\si{\micro\metre}$. $V_\mathrm{DS} = -0.1\,\si{\volt}$. First 20 cycles with $1.2\,\si{\milli\volt\per\second}$. \label{fig:S_SolidState}}
     \vspace*{-\baselineskip}
\end{figure}

\begin{figure}[H]
    \centering
    \includegraphics[width=\linewidth]{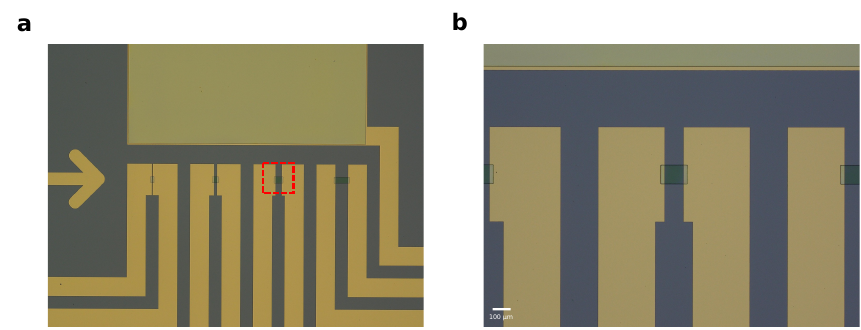}\caption{\textbf{Micrograph of OECT devices.} OECTs with the side-gate architecture used for liquid electrolytes. \label{fig:S_Devices}}
     \vspace*{-\baselineskip}
\end{figure}

\begin{figure}[H]
    \centering
    \includegraphics[width=\linewidth]{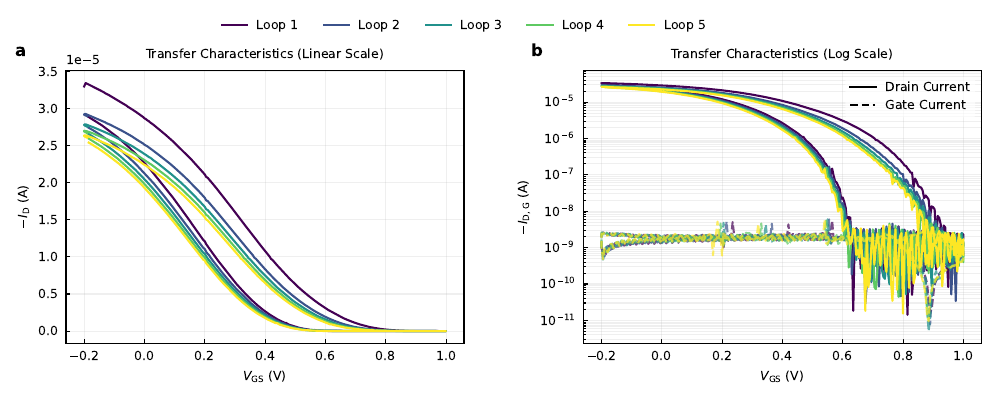}\caption{\textbf{Transfer characteristics for an OECT with liquid \ce{[EMIM][EtSO4]}}. (\textbf{a}) Linear scale, (\textbf{b}) logarithmic scale. As the solid-state system (Fig.\,\ref{fig:S_SolidState}), the OECT features pronounced hysteresis and a low off-current. $V_\mathrm{DS} = -10\,\si{\milli\volt}$. First 5 cycles with $29.1\,\si{\milli\volt\per\second}$. \label{fig:S_Transfer_Cycling}}
     \vspace*{-\baselineskip}
\end{figure}

\begin{figure}[H]
    \centering
    \includegraphics[width=0.4\linewidth]{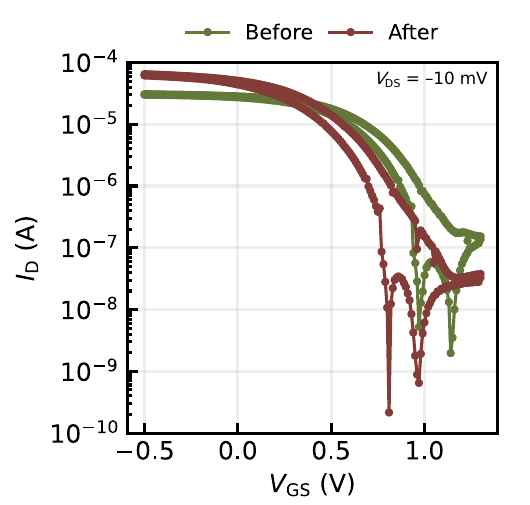} 
    \caption{\textbf{OECT transfer characteristics after \ce{[EMIM][EtSO4]} post-treatment.} An OECT was exposed to \ce{[EMIM][EtSO4]} for $5\,\si{\minute}$, rinsed with deionized water, and measured with 100\,\si{\milli\molar} aqueous \ce{NaCl} electrolyte (before and after exposure, respectively). The treatment reduces the threshold voltage, raises the on-current, and lowers the off-current.\label{fig:S_NaCl_Exposure}}
     \vspace*{-\baselineskip}
\end{figure}

\begin{figure}[H]
    \centering
    \includegraphics[width=\linewidth]{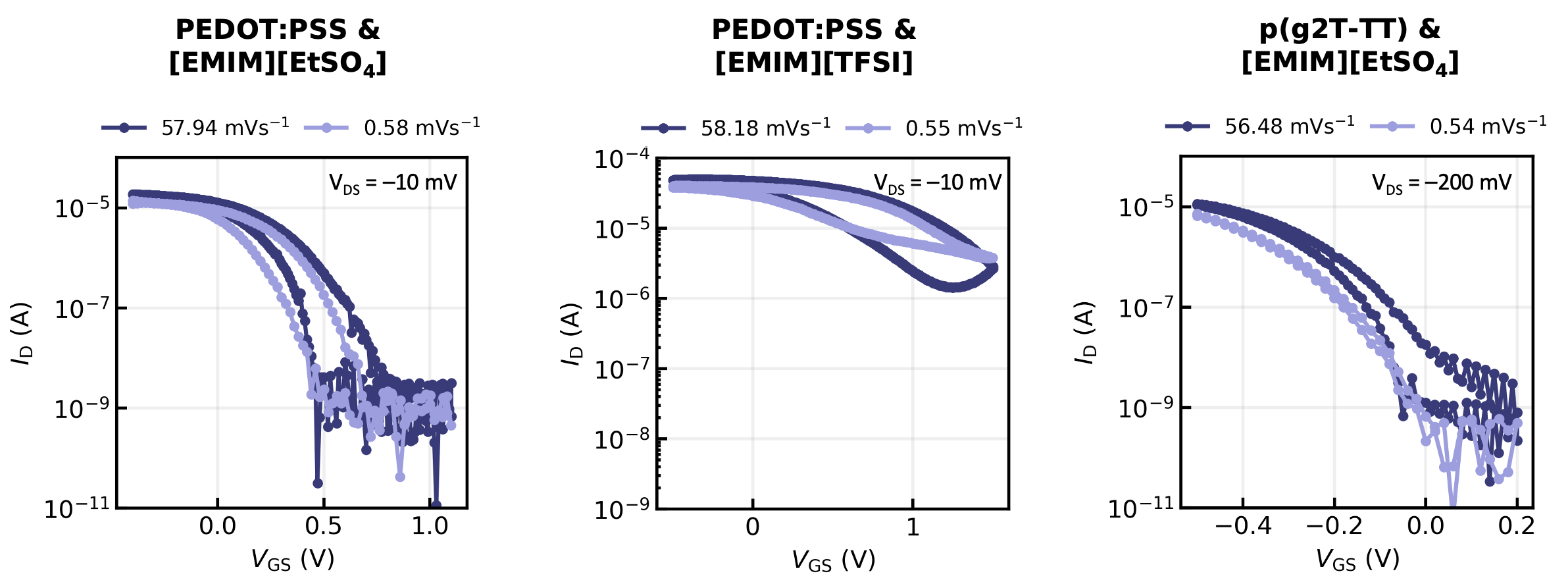} 
    \caption{\textbf{OECT transfer characteristics for different OMIECs and electrolytes.} Transfer characteristics at different scan rates for OECTs based on PEDOT:PSS with \ce{[EMIM][EtSO4]} and \ce{[EMIM][TFSI]} electrolyte, as well as for an OECT with \ce{[EMIM][EtSO4]} electrolyte and a poly(2-(3,3'-bis(2-(2-(2-methoxyethoxy)ethoxy)ethoxy)-[2,2'-bithiophen]-5-yl) thieno [3,2-b]thiophene) (p(g2T-TT)) channel and gate. \label{fig:S_Transfer}}
     \vspace*{-\baselineskip}
\end{figure}

\begin{figure}[H]
    \centering
    \includegraphics[width=\linewidth]{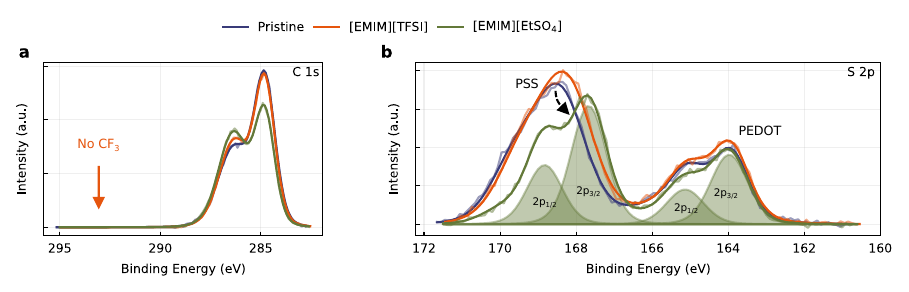} 
    \caption{\textbf{XPS thin-film analysis (C\,1s and S\,2p core levels).} XPS spectra of PEDOT:PSS thin-films exposed to \ce{[EMIM][TFSI]} and \ce{[EMIM][EtSO4]}. (\textbf{a}) C\,1s core level region. No \ce{CF3} signal is detected for the \ce{[EMIM][TFSI]}-treated sample (\ce{CF3} $\sim 293\,\si{\electronvolt}$\autocite{foelske2011xps}). \ce{[EMIM][EtSO4]} induces an altered ratio of \ce{C-O} and \ce{C-C} signal (\ce{C-C} decreases). (\textbf{b}) S\,2p core level region. \ce{[EMIM][TFSI]} does not cause a significant change to the ratio of PEDOT to PSS, aligning with the minute decrease observed in the O\,1s data (Fig.\,\ref{fig:2}c). With \ce{[EMIM][EtSO4]}, on the other hand, the fraction of PEDOT is increased, driven by a removal of PSS. Here, the PSS signal also undergoes a marked shift to lower binding energy. This shift is on the one hand attributed to an overlap with the signal of retained \ce{[EtSO4]-}, having a slightly decreased binding energy compared to the \ce{SO3}-group of PSS\autocite{jurado2016effect}. Moreover, it underlines the binding of the two anionic species to \ce{[EMIM]+}, the $\pi$-system of which strongly increases the vicinal electron density around the sulfur atoms.\label{fig:S_XPS_C_S}}
     \vspace*{-\baselineskip}
\end{figure}

\begin{figure}[H]
    \centering
    \includegraphics[width=\linewidth]{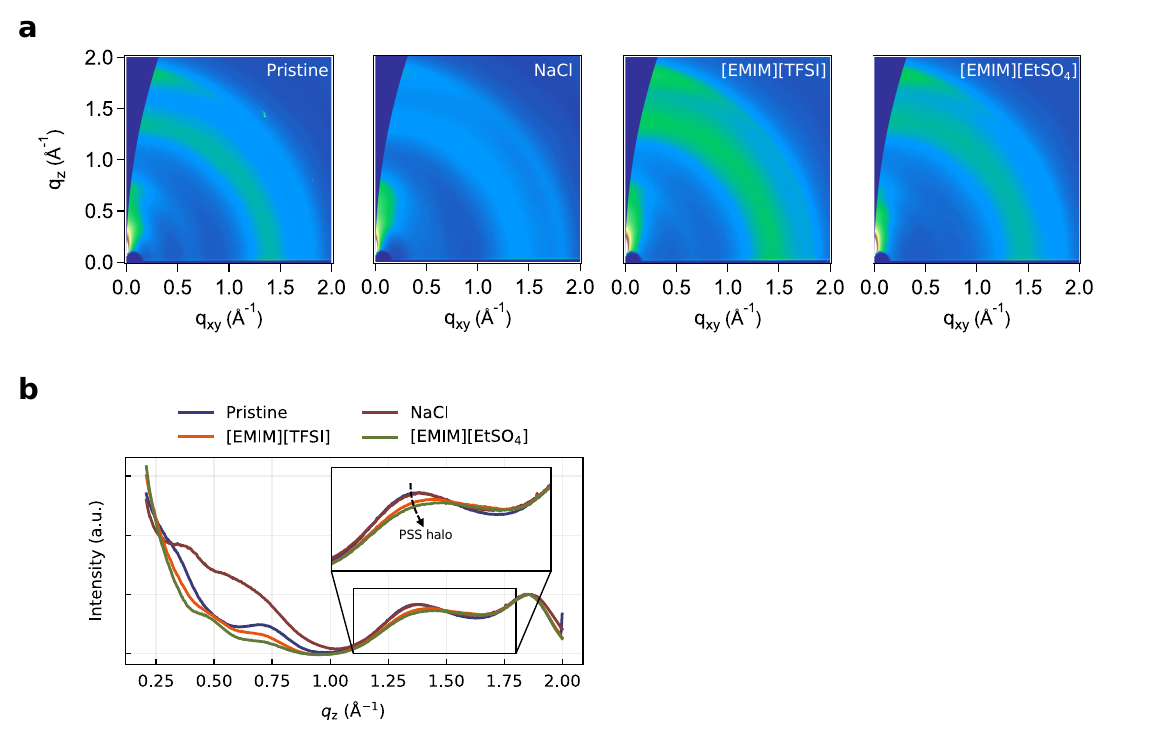} 
    \caption{\textbf{GIWAXS diffraction pattern}. (\textbf{a}) GIWAXS is performed for thin-films of pristine PEDOT:PSS and after exposure to \ce{NaCl} (100\,mM), \ce{[EMIM][TFSI]}, and \ce{[EMIM][EtSO4]}. (\textbf{b}) Normalized GIWAXS profiles. For the pristine film, key features include $(100)$ and $(200)$ lamellar reflections at $q_z\sim0.35\,\si{\per\angstrom}$ and $q_z\sim0.7\,\si{\per\angstrom}$ ($d_{100} = 1.80\,\si{\nano\metre}$), an isotropic PSS halo at $q_z\sim1.38\,\si{\per\angstrom}$, and a PEDOT $\pi$-stacking peak at $q_z=1.85\,\si{\per\angstrom}$ ($d_{010} = 0.34\,\si{\nano\metre}$)\autocite{kim2018influence, li2022deciphering, kee2016controlling}. Exposure to \ce{[EMIM][EtSO4]} induces several changes. The PSS halo broadens and shifts to $q_z=1.44\,\si{\per\angstrom}$, reflecting PSS removal and overlapping reflections from \ce{[EMIM][EtSO4]}\autocite{han2020insight}. The PEDOT $\pi$-stacking peak increases in intensity, indicating enhanced crystallinity, which aligns with previous studies\autocite{kee2016controlling} and resembles structures observed with PEDOT:PSS films post-treated with sulfuric acid\autocite{kim2014highly, kim2018influence}. A new shoulder emerges at $q_z=0.50\,\si{\per\angstrom}$ ($d=1.26\,\si{\nano\metre}$), attributed to the lamellar structure of retained ionic liquid\autocite{han2020insight}, while the lamellar reflections decrease in relative intensity. For the reference systems, \ce{NaCl} introduces minimal changes, with a more intense low-angle region and a minor decrease in $\pi$-stacking distance. \ce{[EMIM][TFSI]} induces similar but less pronounced effects compared to \ce{[EMIM][EtSO4]}, lacking the shoulder at $q_z\sim0.50\,\si{\per\angstrom}$ and showing weaker changes to the PSS halo.
    \label{fig:S_GIWAXS}}
     \vspace*{-\baselineskip}
\end{figure}

\begin{figure}[H]
    \centering
    \includegraphics[width=0.7\linewidth]{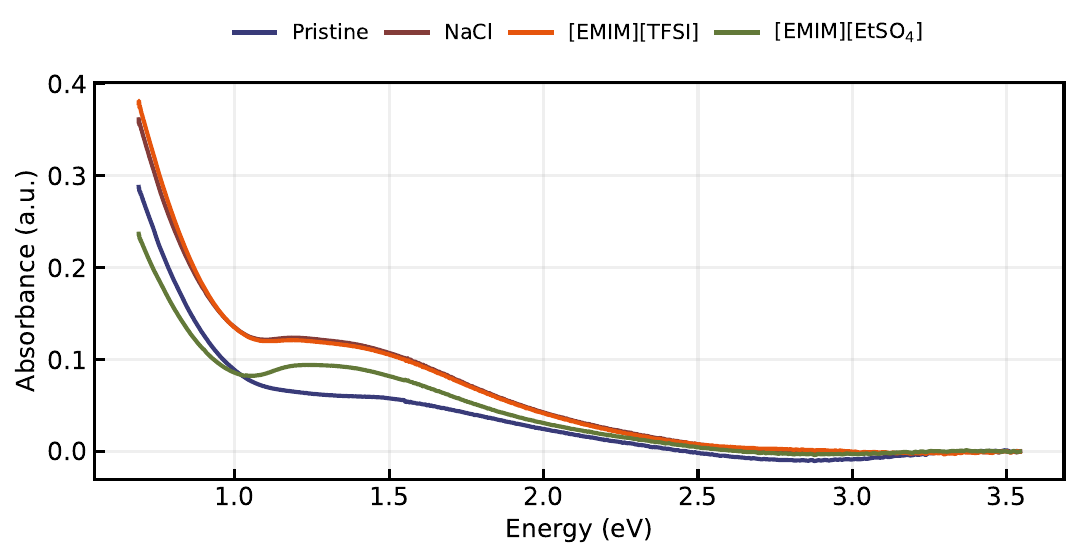} 
    \caption{\textbf{Thin-film UV-Vis spectroscopy}. PEDOT:PSS thin-films were temporarily exposed to the electrolytes and analyzed by UV-Vis spectroscopy. No significant changes to the absorbance are found, except a seemingly more distinct polaron absorbance at around $1.3\,\si{\electronvolt}$ for the sample treated with \ce{[EMIM][EtSO4]}.
    \label{fig:S_UVVis}}
     \vspace*{-\baselineskip}
\end{figure}

\begin{figure}[H]
    \centering
    \includegraphics[width=\linewidth]{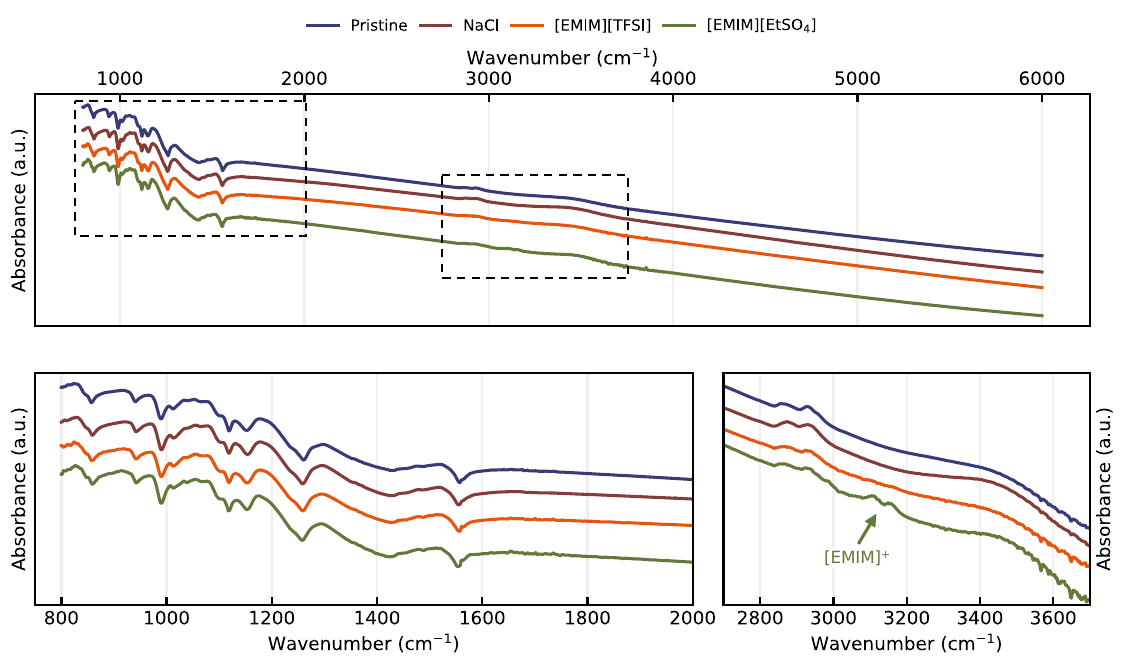} 
    \caption{\textbf{Thin-film IR spectroscopy.} IR spectra recorded for thin-films of pristine and electrolyte-treated PEDOT:PSS. No significant changes are found, except for an additional feature around $3110$--$3160\,\si{\per\centi\metre}$ for the sample treated with \ce{[EMIM][EtSO4]}. The signal can be assigned to the \ce{[EMIM]+} cation retained in the film\autocite{wheeler2017thermal}. Further peak assignment is provided in Tab.\,\ref{tab:S_IR}.
    \label{fig:S_IR}}
     \vspace*{-\baselineskip}
\end{figure}

\begin{figure}[H]
    \centering
    \includegraphics[width=\linewidth]{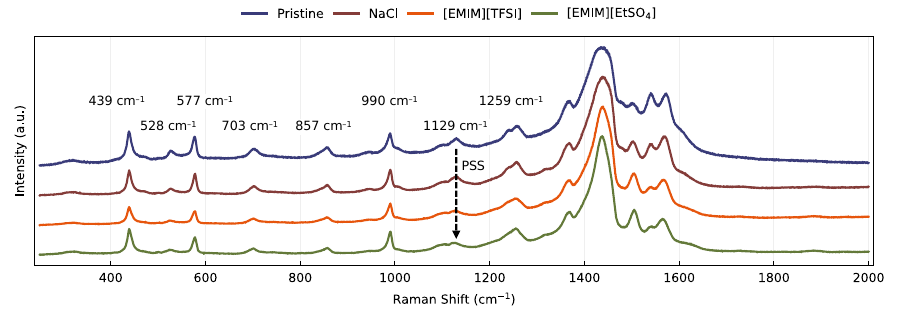} 
    \caption{\textbf{Raman spectroscopy.} Raman spectra measured with $\lambda_\mathrm{exc}=532\,\si{\nano\metre}$. A comprehensive peak assignment is provided in Tab.\,\ref{tab:S_Raman}. The high-energy range is shown isolated in Fig.\,\ref{fig:2}d.
    \label{fig:S_Raman}}
     \vspace*{-\baselineskip}
\end{figure}

\begin{figure}[H]
    \centering
    \includegraphics[width=1\linewidth]{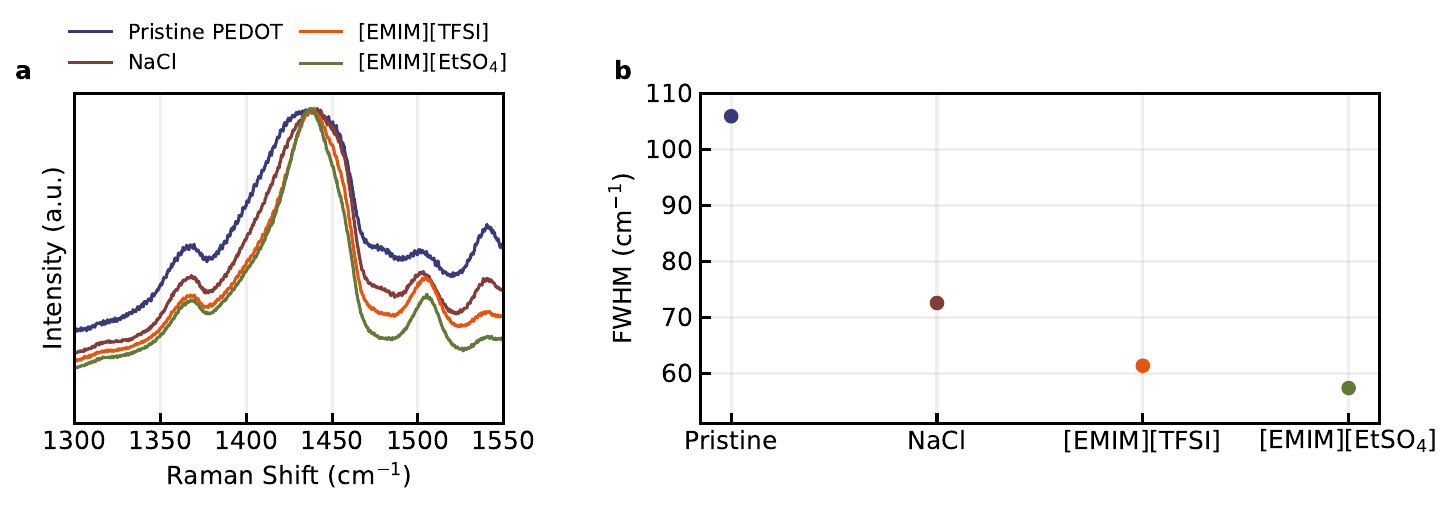} 
    \caption{\textbf{Raman peak analysis.} (\textbf{a}) Principle Raman peaks are identified at $1441\,\si{\per\centi\metre}$ (pristine), $1439\,\si{\per\centi\metre}$ (NaCl), $1439\,\si{\per\centi\metre}$ (\ce{[EMIM][TFSI]}), and $1438\,\si{\per\centi\metre}$ (\ce{[EMIM][EtSO4]}). (\textbf{b}) Electrolyte exposure effects a reduction in bandwidth of the principle peak, evident in a decreasing full width at half maximum (FWHM).\label{fig:S_Raman_FWHM}}
     \vspace*{-\baselineskip}
\end{figure}

\begin{figure}[H]
    \centering
    \includegraphics[width=0.9\linewidth]{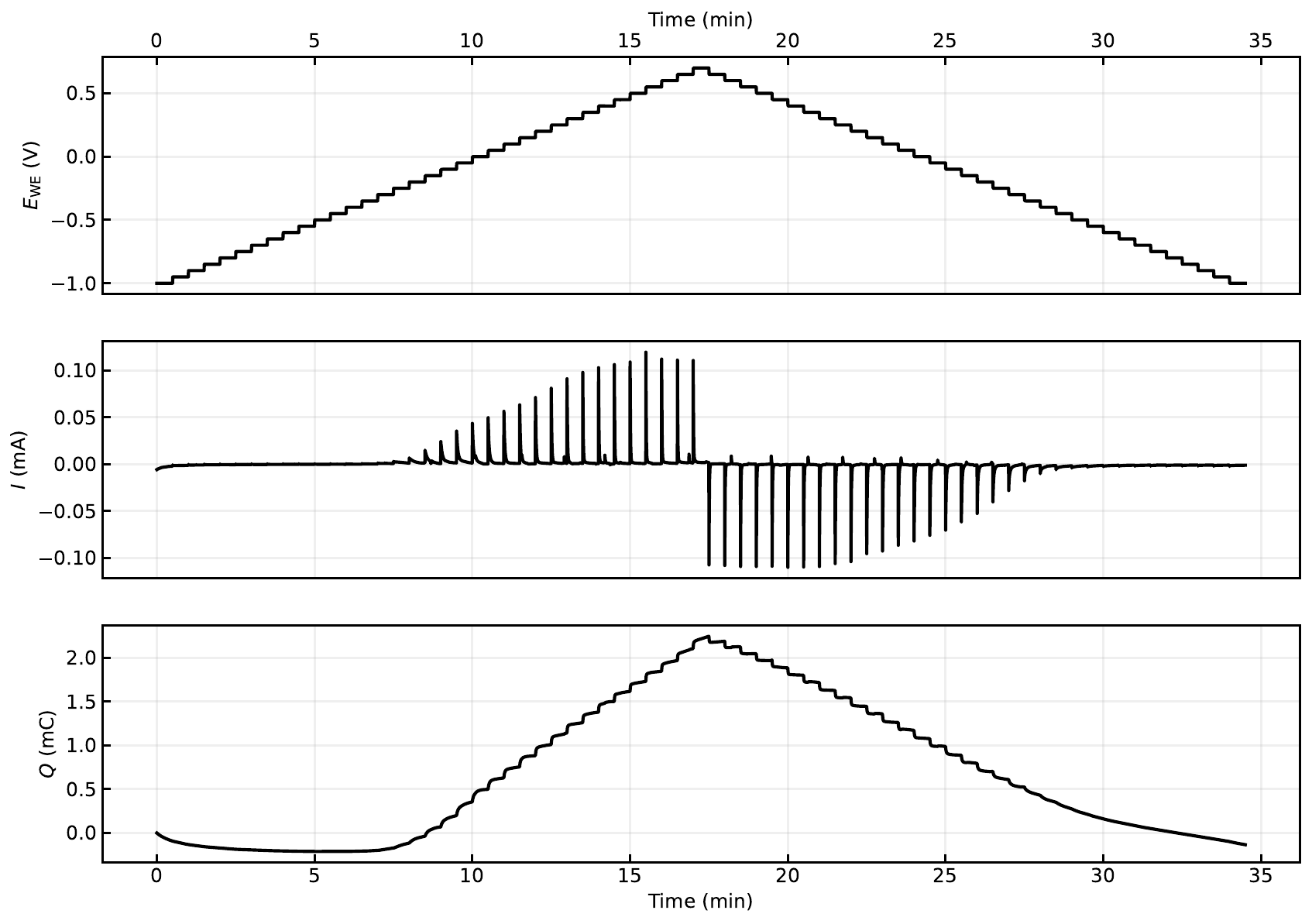} 
    \caption{\textbf{Chronoamperometry data for PEDOT:PSS with \ce{[EMIM][EtSO4]}.} The potential at the working electrode ($E_\mathrm{WE}$) was modulated in increments of $50\,\si{\milli\volt}$ between $-1.0\,\si{\volt}$ and $0.7\,\si{\volt}$, holding each potential for $30\,\si{\second}$.
    \label{fig:S_CA}}
     \vspace*{-\baselineskip}
\end{figure}

\begin{figure}[H]
    \centering
    \includegraphics[width=\linewidth]{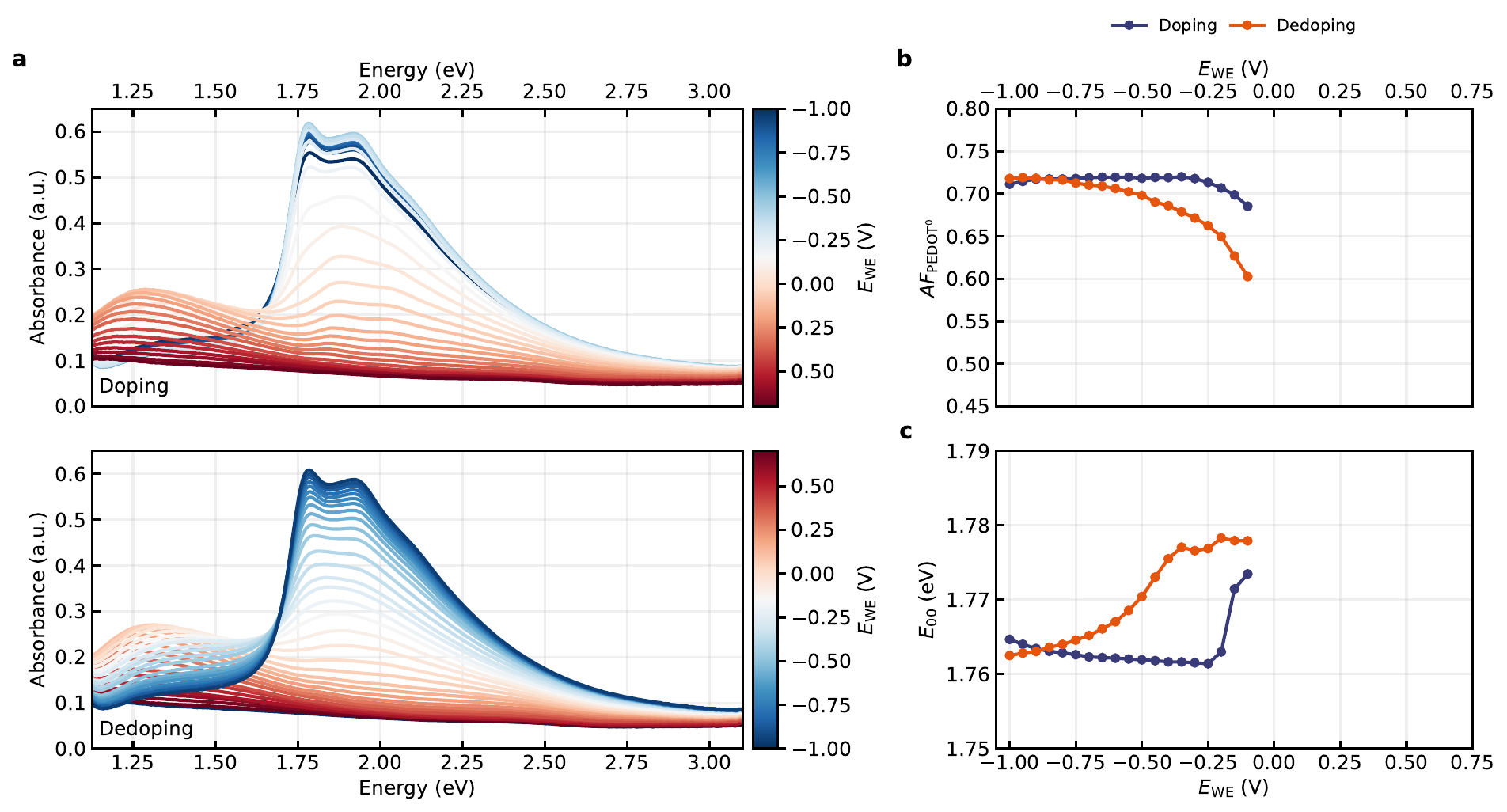} 
    \caption{\textbf{Spectroelectrochemical measurement with \ce{[EMIM][EtSO4]}}. (\textbf{a}) Absorbance spectra recorded under doping and dedoping. (\textbf{b}) Aggregate fraction and (\textbf{c}) $E_{00}$ energies under doping and dedoping. The exceeding intensity of the 0--0 transition compared to the 0--1 transition indicates a dominance of intrachain coupling.
    \label{fig:S_Spec_EMIMEtSO4}}
     \vspace*{-\baselineskip}
\end{figure}

\begin{figure}[H]
    \centering
    \includegraphics[width=\linewidth]{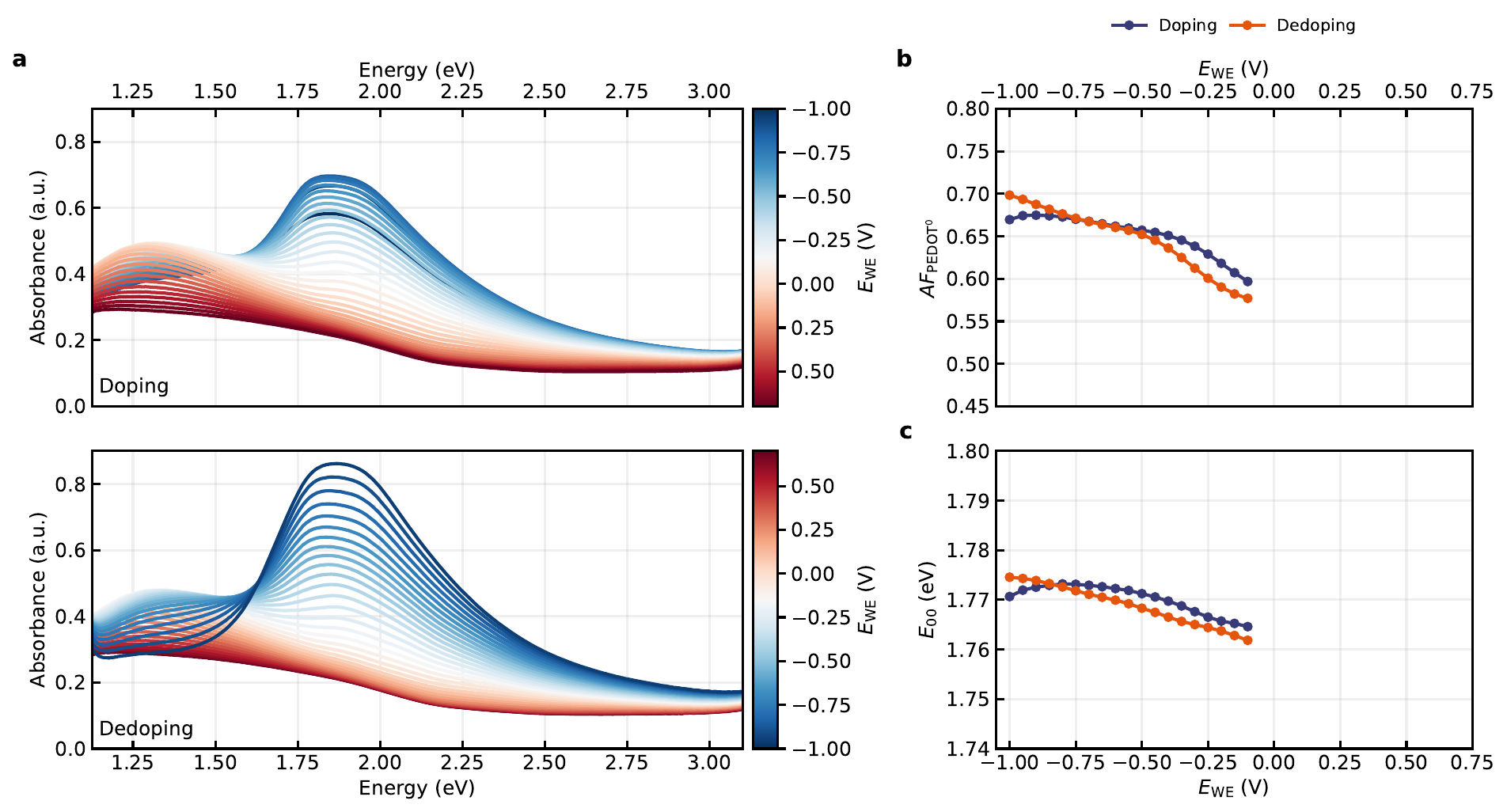} 
    \caption{\textbf{Spectroelectrochemical measurement with \ce{[EMIM][TFSI]}.}
(\textbf{a}) Absorbance spectra recorded under doping and dedoping. (\textbf{b}) Aggregate fraction and (\textbf{c}) $E_{00}$ energies under doping and dedoping.\label{fig:S_Spec_EMIMTFSI}}
     \vspace*{-\baselineskip}
\end{figure}

\begin{figure}[H]
    \centering
    \includegraphics[width=\linewidth]{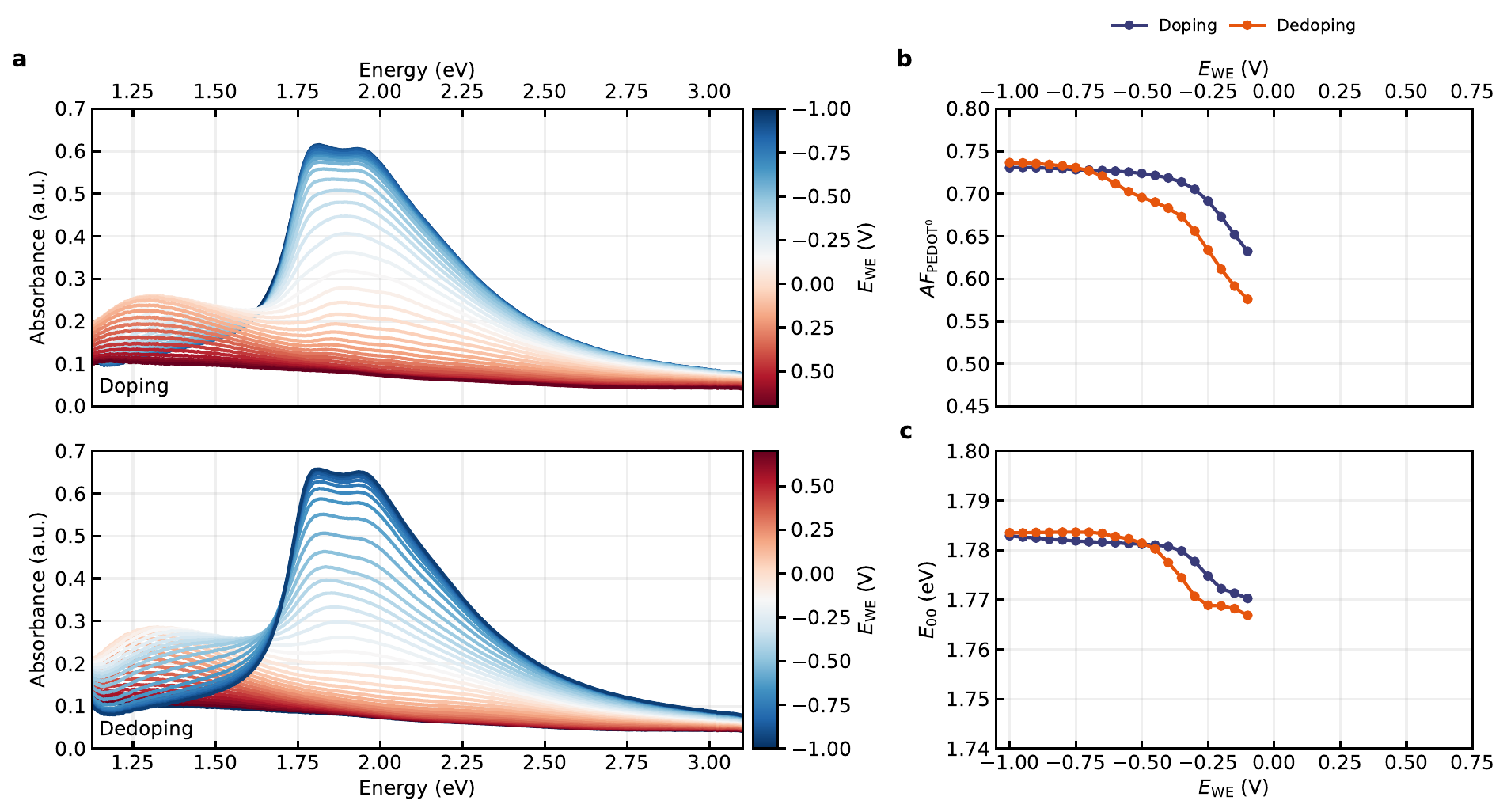} 
    \caption{\textbf{Spectroelectrochemical measurement with \ce{NaCl} (100\,mM).}  (\textbf{a}) Absorbance spectra recorded under doping and dedoping. (\textbf{b}) Aggregate fraction and (\textbf{c}) $E_{00}$ energies under doping and dedoping. The balancing intensity of the 0--0 transition compared to the 0--1 transition indicates a balance of intra- and interchain coupling.    \label{fig:S_Spec_NaCl}}
     \vspace*{-\baselineskip}
\end{figure}

\begin{figure}[H]
    \centering
    \includegraphics[width=0.8\linewidth]{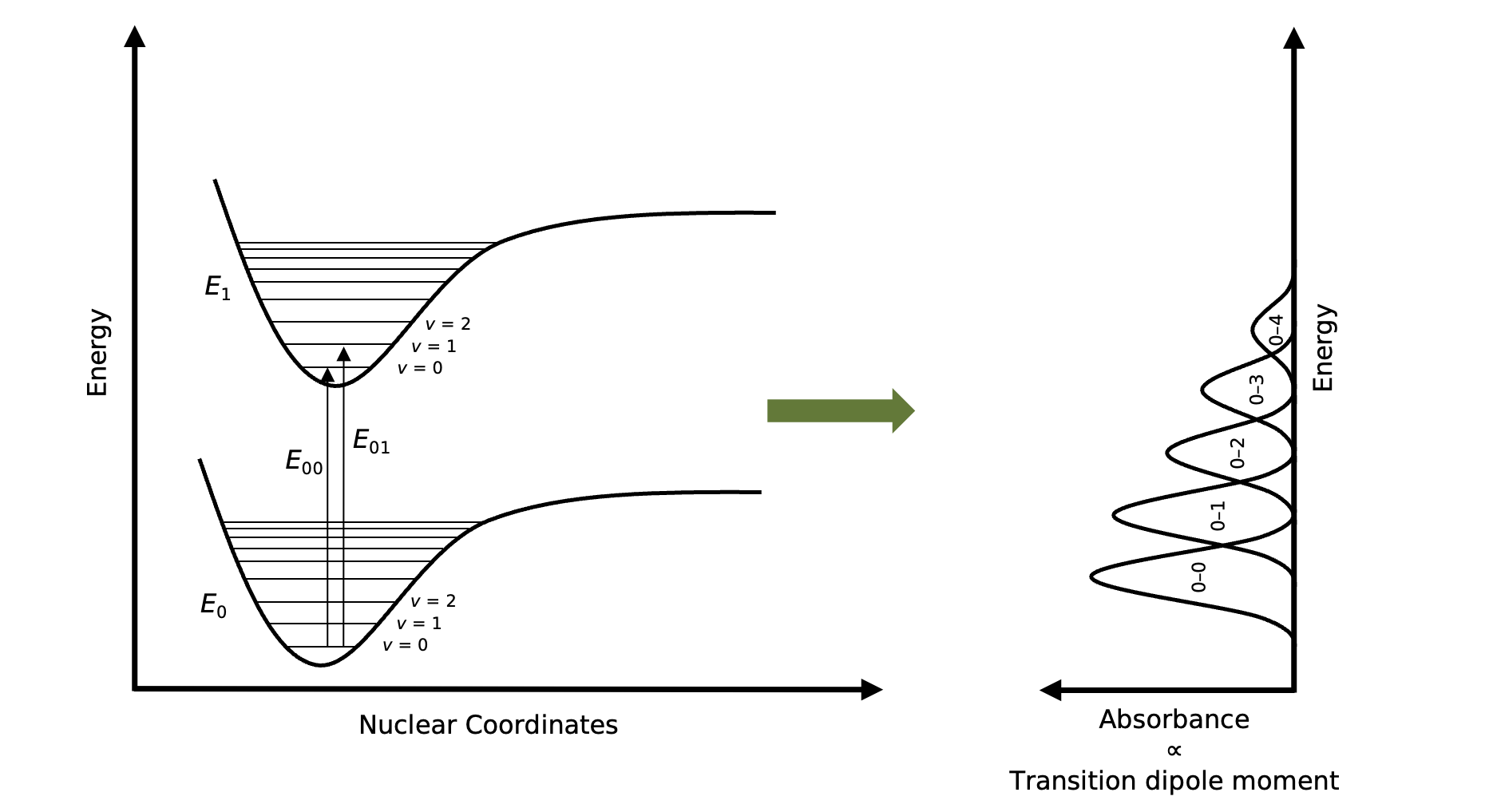} 
    \caption{\textbf{Vibronic transitions.} Schematic illustration of the electronic and vibrational transitions underlying the absorbance spectra in spectroelectrochemistry. Electronic excitations go along with subordinate vibrational transitions. The energetic difference between the ground ($E_0$) and excited ($E_1$) state can be traced by the absorbance line $E_{00}$, with smaller values (red-shift) indicating thermodynamically stabilized species. \label{fig:S_Transition}}
     \vspace*{-\baselineskip}
\end{figure}

\begin{figure}[H]
    \centering
    \includegraphics[width=\linewidth]{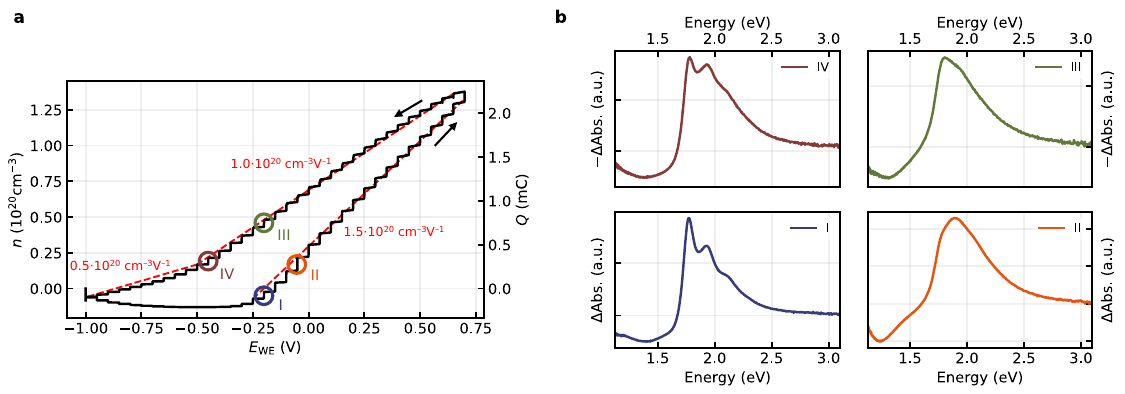} 
    \caption{\textbf{De-/doping cycle of PEDOT:PSS and differential absorbance.} (\textbf{a}) Charge carrier density ($n$) as a function of applied potential $E_\mathrm{WE}$ (V vs. Ag/AgCl). Different charging efficiencies (differential capacitances) are found for charging and discharging. The latter shows a flattening kink at around $E_\mathrm{WE}=-0.5\,\si{\volt}$, indicating energetic impediment. (\textbf{b}) Differential absorbance spectra corresponding to the four potential steps highlighted in (\textbf{a}). Ordered species are charged first, followed by disordered, more amorphous entities. Discharging the former is offset to more negative potentials, indicating an energetic stabilization of the doped state. Note also the difference in relative peak intensities between I (doping) and IV (dedoping), indicative of a disruption of the local electronic environment.
    \label{fig:S_CA_Diff_EMIMEtSO4}}
     \vspace*{-\baselineskip}
\end{figure}

\begin{figure}[H]
    \centering
    \includegraphics[width=\linewidth]{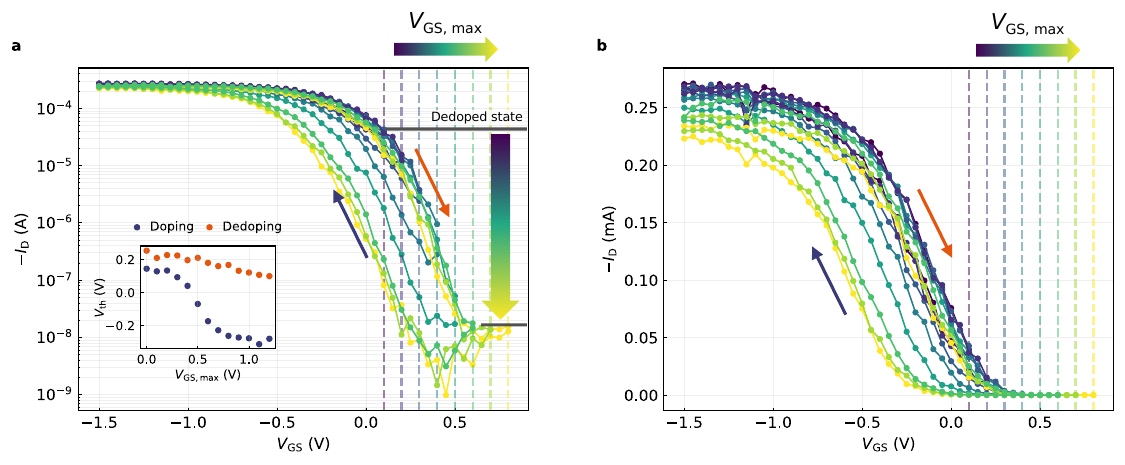} 
    \caption{\textbf{Dependence of hysteresis on scan range.} Transfer characteristics of a PEDOT:PSS-based OECT with an \ce{[EMIM][EtSO4]} solid‐state electrolyte as reported in ref.\,\citen{weissbach2022photopatternable}. (\textbf{a}) Logarithmic scale, (\textbf{b}) linear scale scale. Inset: threshold voltages $V_\mathrm{th}$ for the doping and dedoping branches, extracted by linear fits of the transfer curves ($V_\mathrm{DS}=-0.1\,\si{\volt}$). Each sweep begins in the fully doped state at $V_\mathrm{GS}=-1.5\,\si{\volt}$ (depletion mode). By raising the maximum gate voltage, the channel is driven to progressively deeper dedoped states, such that (re‐)doping is probed from different starting conditions.
    \\ \\
    For small dedoping, the two branches coincide ($V_\mathrm{th}\approx0.2\,\si{\volt}$) and hysteresis is negligible. As the dedoping limit is increased, the doping branch shifts to lower voltages, revealing distinct energetics: from the most strongly dedoped state, doping takes place with a threshold voltage lowered by around $400\,\si{\milli\volt}$ as compared to dedoping.
    \\ \\
    This observation agrees with our model laid out in the main text: \ce{[EtSO4]-} anions not only maintain charge neutrality with the polarons on \ce{PEDOT+}, but also bind strongly to those charged PEDOT chains, thus stabilizing the doped state. In a strongly dedoped channel, only a few anions remain, likely in aggregated domains that create deep hole‐stabilizing sites. These sites are filled first during doping and emptied last during dedoping, maximizing hysteresis when reaching the most depleted state. By contrast, partial dedoping removes only carriers from higher‐energy, amorphous regions---where the conventional \ce{PEDOT+}/\ce{PSS-} compensation is more reversible---resulting in minimal hysteresis when cycling is restricted to this range.}
    \label{fig:S_ScanRange}
    \vspace*{-\baselineskip}
\end{figure}

\addcontentsline{toc}{subsection}{Fig.~\ref{fig:S_PEDOTPSS}\quad Molecular structure of PEDOT:PSS}
\addcontentsline{toc}{subsection}{Fig.~\ref{fig:S_SolidState}\quad Transfer characteristics for a bistable, solid-state OECT}
\addcontentsline{toc}{subsection}{Fig.~\ref{fig:S_Devices}\quad Micrograph of OECT devices}
\addcontentsline{toc}{subsection}{Fig.~\ref{fig:S_Transfer_Cycling}\quad Transfer characteristics for an OECT with liquid \ce{[EMIM][EtSO4]}}
\addcontentsline{toc}{subsection}{Fig.~\ref{fig:S_NaCl_Exposure}\quad OECT transfer characteristics after \ce{[EMIM][EtSO4]} post-treatment}
\addcontentsline{toc}{subsection}{Fig.~\ref{fig:S_Transfer}\quad OECT transfer characteristics for different OMIECs and electrolytes}
\addcontentsline{toc}{subsection}{Fig.~\ref{fig:S_XPS_C_S}\quad XPS thin-film analysis (C\,1s and S\,2p core levels)}
\addcontentsline{toc}{subsection}{Fig.~\ref{fig:S_GIWAXS}\quad GIWAXS diffraction pattern}
\addcontentsline{toc}{subsection}{Fig.~\ref{fig:S_UVVis}\quad Thin-film UV-Vis spectroscopy}
\addcontentsline{toc}{subsection}{Fig.~\ref{fig:S_IR}\quad Thin-film IR spectroscopy}
\addcontentsline{toc}{subsection}{Fig.~\ref{fig:S_Raman}\quad Raman spectroscopy
}
\addcontentsline{toc}{subsection}{Fig.~\ref{fig:S_Raman_FWHM}\quad Raman peak analysis
}
\addcontentsline{toc}{subsection}{Fig.~\ref{fig:S_CA}\quad Chronoamperometry data for PEDOT:PSS with \ce{[EMIM][EtSO4]}
}
\addcontentsline{toc}{subsection}{Fig.~\ref{fig:S_Spec_EMIMEtSO4}\quad Spectroelectrochemical measurement with \ce{[EMIM][EtSO4]}}
\addcontentsline{toc}{subsection}{Fig.~\ref{fig:S_Spec_EMIMTFSI}\quad Spectroelectrochemical measurement with \ce{[EMIM][TFSI]}}
\addcontentsline{toc}{subsection}{Fig.~\ref{fig:S_Spec_NaCl}\quad Spectroelectrochemical measurement with \ce{NaCl}}
\addcontentsline{toc}{subsection}{Fig.~\ref{fig:S_Transition}\quad Vibronic transitions in spectroelectrochemistry}
\addcontentsline{toc}{subsection}{Fig.~\ref{fig:S_CA_Diff_EMIMEtSO4}\quad De-/doping cycle of PEDOT:PSS and differential absorbance}
\addcontentsline{toc}{subsection}{Fig.~\ref{fig:S_ScanRange}\quad Dependence of hysteresis on scan range}
\addcontentsline{toc}{subsection}{Fig.~\ref{fig:S_Mobility}\quad Mobility measurements}
\addcontentsline{toc}{subsection}{Fig.~\ref{fig:S_XPS_S}\quad XPS spectra (S\,2p) of PEDOT:PSS for molar ratio estimation}
\addcontentsline{toc}{subsection}{Fig.~\ref{fig:S_XPS_O}\quad XPS spectra (O\,1s) of PEDOT:PSS for ion exchange estimation}
\addcontentsline{toc}{subsection}{Fig.~\ref{fig:S_Composition}\quad Composition of PEDOT:PSS channels after exposure to ionic liquids}
\addcontentsline{toc}{subsection}{Fig.~\ref{fig:S_fit_results}\quad Representative vibronic fits}
\addcontentsline{toc}{subsection}{Fig.~\ref{fig:E_00_note}\quad Ground state transition energies from manual fits}

\addcontentsline{toc}{section}{Supplementary Tables}

\addcontentsline{toc}{subsection}{Table~\ref{tab:S_IR}\quad IR peak assignment}
\addcontentsline{toc}{subsection}{Table~\ref{tab:S_Raman}\quad Raman peak assignment}
\addcontentsline{toc}{subsection}{Table~\ref{tab:S_XPS_Ratios}\quad PEDOT:PSS molar ratios estimated from XPS}
\addcontentsline{toc}{subsection}{Table~\ref{tab:O_XPS_Ratios}\quad Ion-exchange ratios estimated from XPS}

\nocite{dijk2023pedot, weissbach2022photopatternable, bongartz2024bistable}

\nocite{karri2019synthesis, petsagkourakis2023improved}
\begin{table}[h!]
\caption{\textbf{IR peak assignment.}}
\centering
\begin{tabular}{r r l r}
\hline
Wavenumber ($\si{\per\centi\metre}$) & Lit. ($\si{\per\centi\metre}$)& Mode & Ref.\\
\hline
857, 940 & 842, 955&  \ce{C-S-C} & \citen{petsagkourakis2023improved}\\
989, 1119 & 1055, 1126 & \ce{C-O-C} & \citen{petsagkourakis2023improved} \\
1013 & 1026 & \ce{C-S} (PSS)  & \citen{petsagkourakis2023improved}\\
1153 & 1172 & \ce{S=O} (sym.) & \citen{petsagkourakis2023improved} \\
1261 & 1273 & \ce{C-C}, \ce{S=O} (asym.) & \citen{petsagkourakis2023improved} \\
1429 & 1412--1418& \ce{C=C} (sym.) & \citen{petsagkourakis2023improved} \\
1557 & 1540--1550& \ce{C=C} (asym.) & \citen{petsagkourakis2023improved} \\
2866, 2932 & 2851, 2925 & \ce{C-H}  & \citen{karri2019synthesis}\\
$\sim3440$ &$\sim3450$  &   \ce{O-H} & \citen{karri2019synthesis}\\ \hline
3110, 3155 &  3103, 3151 &   \ce{C-H} (\ce{[EMIM]}) & \citen{wheeler2017thermal}\\
\hline
\end{tabular}
\label{tab:S_IR}
\end{table}

\begin{table}[h!]
\caption{\textbf{Raman peak assignment.} Literature values correspond to the same excitation wavelength ($\lambda_\mathrm{exc}=532\,\si{\nano\metre}$).}
\centering
\begin{tabular}{r r l}
\hline
Raman Shift ($\si{\per\centi\metre}$) & Lit.\autocite{jucius2019structure} ($\si{\per\centi\metre}$)& Mode \\
\hline
439 & 439 &  \\
528 &  $\sim525$ &   \\
577 & 577 & Oxyethylene ring deform.-vibr.  \\
703 & 702 & C-S-C (sym.) deform.   \\
857 &  855 & Oxyethylene ring deform.-vibr. (C-H bending)  \\
990 &  988 & Oxyethylene ring deform.-vibr. \\
1129 &  1124 & PSS   \\
1259 &  1256 &  \ce{C_\alpha-C_{\alpha'}} inter-ring  stretch.-vibr. \\
1367 &  1367&  \ce{C_\beta-C_\beta} stretch.-deform.  \\
1438 & 1428 & \ce{C_\alpha=C_\beta} (sym.) stretch.-vibr.\\
1502 & 1486 &  \ce{C_\alpha=C_\beta} (asym.) stretch.-vibr. (mid-chain)\\
1540 & 1537 & \ce{C_\alpha=C_\beta} (asym.) stretch.-vibr. (split)\\
1572 & 1569 & \ce{C_\alpha=C_\beta} (asym.) stretch.-vibr. (end-chain) \\

\hline
\end{tabular}
\label{tab:S_Raman}
\end{table}

\nocite{weissbach2022photopatternable}

\let\oldcite\cite
\renewcommand{\cite}[1]{\textsuperscript{\oldcite{#1}}}
\newcommand{\normcite}[1]{\textnormal{\oldcite{#1}}}

\clearpage
\section*{Supplementary Notes}
\addcontentsline{toc}{section}{Supplementary Notes}

\section*{Supplementary Note 1: Mobility Measurements}
\addcontentsline{toc}{subsection}{Supplementary Note 1: Mobility Measurements}
\label{Note_S:Mobility}

We study the hole mobility in OECTs with the \ce{[EMIM][EtSO4]} electrolyte using a previously reported pulsed current technique\autocite{bernards2007steady, keene2022efficient}. We perform measurements before and after biasing the device at $V_\mathrm{GS}=1\,\si{\volt}$ for $60\,\si{\second}$ to probe the doped and undoped state of the channel.

\begin{figure}[H]
    \centering
    \includegraphics[width=\linewidth]{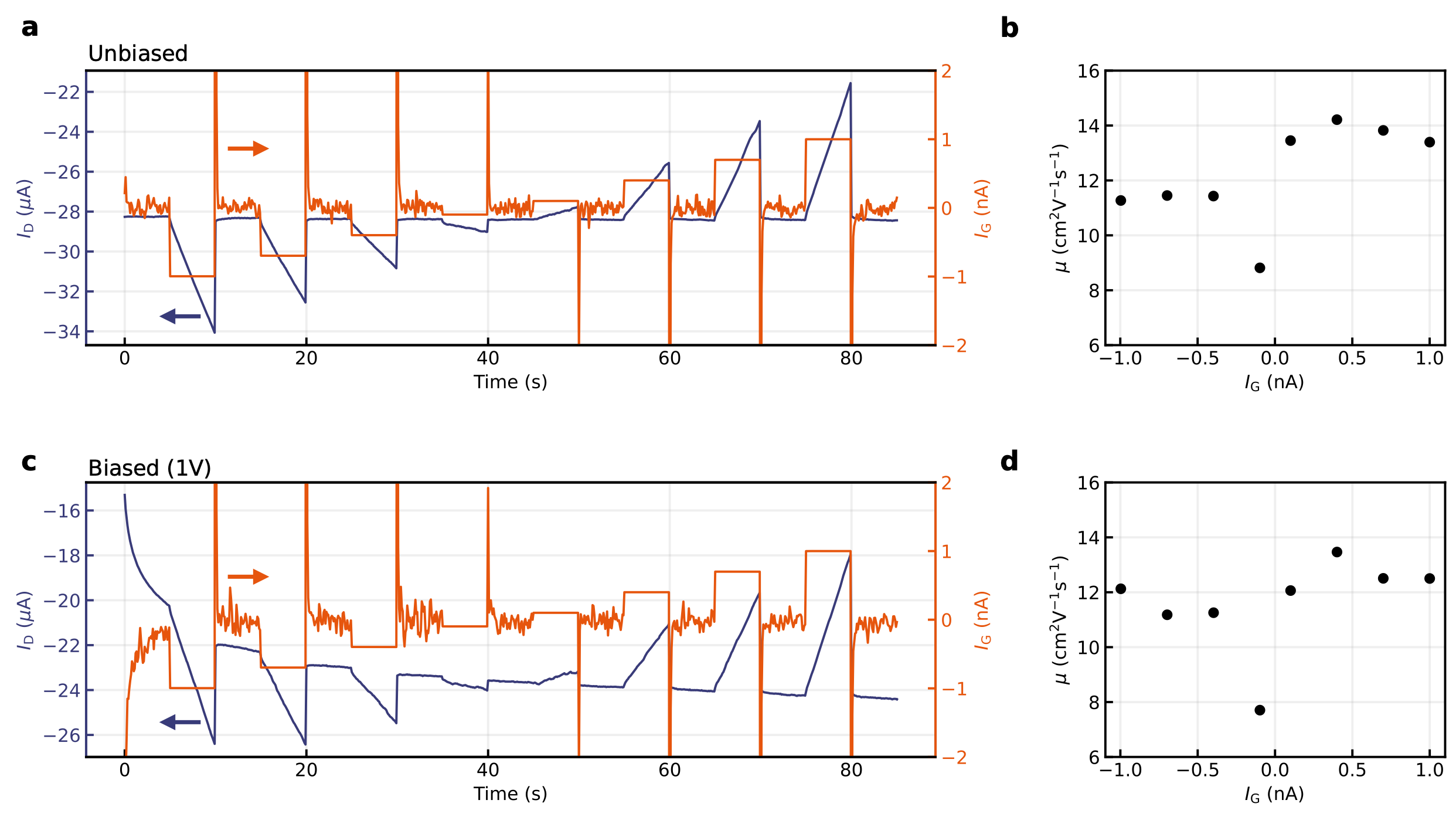} 
    \caption{\textbf{Mobility measurements}. Hole mobilities are measured for an OECT with PEDOT:PSS channel and \ce{[EMIM][EtSO4]} electrolyte by a previously reported pulsed current technique\autocite{bernards2007steady, keene2022efficient}. (\textbf{a}, \textbf{b}) Before biasing and (\textbf{c}, \textbf{d}) after biasing the OECT at $V_\mathrm{GS}=1\,\si{\volt}$ for $60\,\si{\second}$ to measure the doped and undoped state.\label{fig:S_Mobility}}
     \vspace*{-\baselineskip}
\end{figure}
    
\noindent The drain current $I_\mathrm{D}$ is recorded for gate current pulses $I_\mathrm{G}$. From this, the mobility follows as
    \begin{equation}
        \mu = \frac{\partial I_\mathrm{D}}{\partial t} \cdot \frac{L^2}{V_\mathrm{DS}\cdot I_\mathrm{G}}
    \end{equation}
for the individual current pulse, where $L=100\,\si{\micro\metre}$ and $V_\mathrm{DS}=-0.01\,\si{\volt}$. 8 pulses of increasing $I_\mathrm{G}$ are applied for $5\,\si{\second}$ each, yielding mobilities as shown in Fig.\,\ref{fig:S_Mobility}b and d, representing the doped and undoped state of the channel. The averaged mobilities result as 
\begin{align}
    \mu_d & = (12.2\pm0.6)\,\si{\square\centi\metre\per\volt\per\second} \quad \text{and}\nonumber \\
    \mu_u & = (11.6\pm0.6)\,\si{\square\centi\metre\per\volt\per\second}, \nonumber 
\end{align}
with uncertainties corresponding to the standard error of the mean.

\clearpage

\section*{Supplementary Note 2: Ion Exchange Estimations}
\addcontentsline{toc}{subsection}{Supplementary Note 2: Ion Exchange Estimations}
\label{Note_S:Ion_Transfer}

Using DFT, de Izarra et al. calculated the ion exchange free energy for PEDOT:PSS and \ce{[EMIM][EtSO4]} as $1.9\,\si{\kilo\joule\per\mol}$, equivalent to $19.7\,\si{\milli\electronvolt}$\autocite{de2018ionic}. Given a Boltzmann distribution, the exchange ratio follows as
\begin{equation}
    \frac{N_2}{N_1} = e^{-\frac{\Delta G}{k_\mathrm{B}T}},
\end{equation}
which suggests that at room temperature ($298.15\,\si{\kelvin}$), approximately $31.7\%$ of the entities are in the ion-transferred state.

\section*{Supplementary Note 3: XPS Ratio Estimations}
\addcontentsline{toc}{subsection}{Supplementary Note 3: XPS Ratio Estimations}
\label{Note_S:XPS_Ratios}

\begin{figure}[H]
    \centering
    \includegraphics[width=\linewidth]{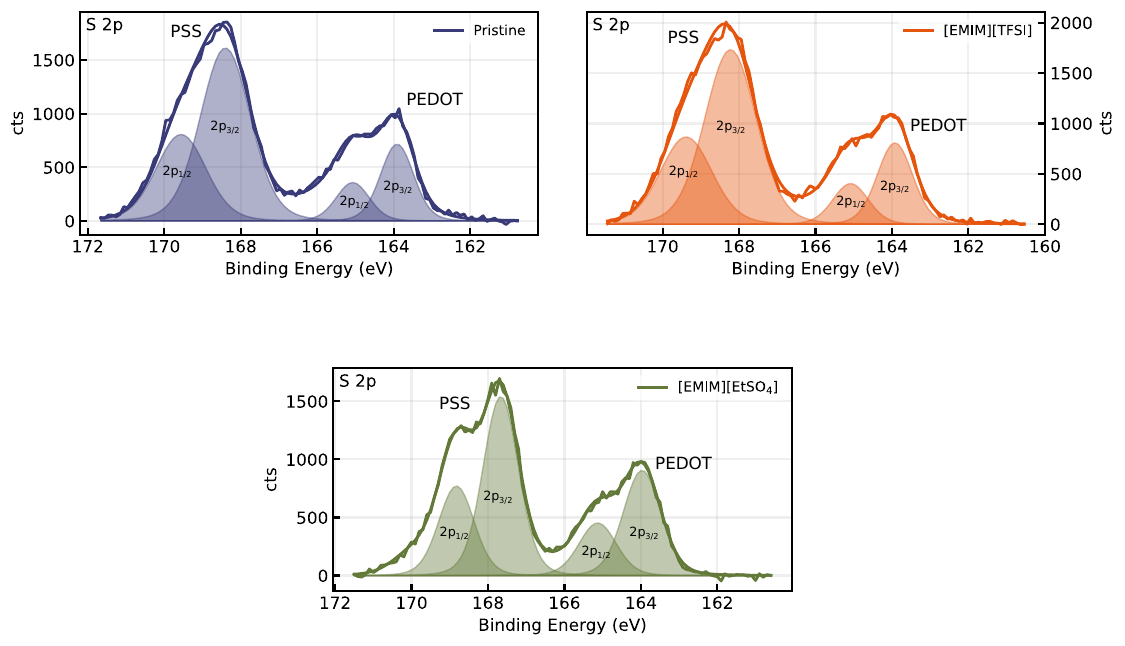} 
    \caption{\textbf{XPS spectra (S\,2p) of PEDOT:PSS for molar ratio estimation.} XPS (S\,2p) spectra and fits of pristine PEDOT:PSS and after exposure to \ce{[EMIM][TFSI]} and \ce{[EMIM][EtSO4]}. The integrated fits serve to estimate the ratio of PEDOT to PSS. \label{fig:S_XPS_S}}
     \vspace*{-\baselineskip}
\end{figure}

We calculate PEDOT-to-PSS (molar) ratios from the XPS data, respectively the fits of the same (Fig.\,\ref{fig:S_XPS_S}). For this, spectra are background-subtracted using a Shirley-type background and smoothed by means of a Savitzky-Golay filter. Peaks are fitted with Gaussian-Lorentzian lineshapes. PEDOT:PSS ratios are inferred from the integrated 2p$_{1/2}$ and 2p$_{3/2}$ signals of PEDOT and PSS, where we assume similar cross-sections for photoelectron emission and therefore similar sensitivities: 

\begin{align}
    \mathrm{cts}^{\mathrm{S\,2p}}(\mathrm{PEDOT}) & = \int{ \mathrm{S\,2p}_{ 2\mathrm{p}_{1/2}}(\mathrm{PEDOT}) \,\mathrm{d}E}+ \int{ \mathrm{S\,2p}_{ 2\mathrm{p}_{3/2}}(\mathrm{PEDOT}) \,\mathrm{d}E} \\ 
    \mathrm{cts}^{\mathrm{S\,2p}}(\mathrm{PSS}) & = \int{ \mathrm{S\,2p}_{ 2\mathrm{p}_{1/2}}(\mathrm{PSS}) \,\mathrm{d}E}+ \int{ \mathrm{S\,2p}_{ 2\mathrm{p}_{3/2}}(\mathrm{PSS}) \,\mathrm{d}E}
\end{align}

\begin{table}[h!]
\caption{\textbf{PEDOT:PSS molar ratios estimated from XPS.}}
\centering
\begin{tabular}{r l l l}
\hline
System &  $\mathrm{cts}^{\mathrm{S\,2p}}(\mathrm{PEDOT})$ & $\mathrm{cts}^{\mathrm{S\,2p}}(\mathrm{PSS})$ &$R_\mathrm{M}$ (PEDOT:PSS)\\
\hline
Pristine & $1217.35$ &  $4125.70$ & $3.39$ \\
\ce{[EMIM][TFSI]} & $1471.60$ &  $4563.76$ & $3.10$ \\
\ce{[EMIM][EtSO4]} & $1737.54$ &  $2803.69$ & $1.61$\\
\hline
\end{tabular}
\label{tab:S_XPS_Ratios}
\end{table}

\begin{figure}[H]
    \centering
    \includegraphics[width=\linewidth]{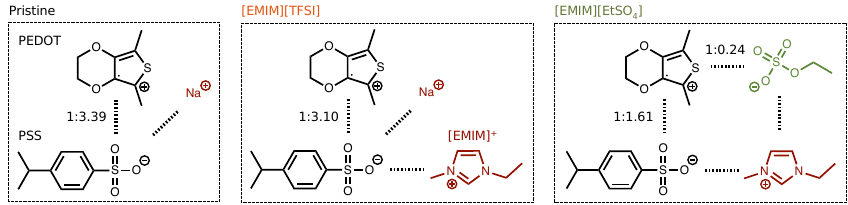} 
    \caption{\textbf{Composition of PEDOT:PSS channels after exposure to ionic liquids.} Formation of a quaternary ionic system with \ce{[EMIM][EtSO4]}, consistent with previous reports\autocite{kee2016controlling}. \label{fig:S_Composition}}
     \vspace*{-\baselineskip}
\end{figure}

\noindent Our calculations show that \ce{[EMIM][EtSO4]} considerably reduces the relative amount of PSS compared to the pristine sample. This process goes along with an ion exchange, where residual \ce{Na+} is replaced by \ce{[EMIM]+} in addition to \ce{[EMIM][EtSO4]} being retained, giving rise to the composition illustrated in Fig.\,\ref{fig:2}f, \ref{fig:S_Composition} and aligning with reports by Kee et al.\autocite{kee2016controlling}.
\\ \\
\noindent We further estimate the degree of ion exchange from the XPS data and in particular, from the O\,1s core level region (Fig.\,\ref{fig:S_XPS_O}). The signal of the pristine sample can be attributed solely to \ce{PSS}, while the \ce{[EMIM][EtSO4]}-treated sample shows signals for both \ce{PSS} and \ce{[EtSO4]-}. To relate the data sets to the same basis (\ce{PEDOT}), we use the molar ratios of Tab.\,\ref{tab:S_XPS_Ratios} and calculate the ratio of ion exchange as follows:

\begin{figure}[t!]
    \centering
    \includegraphics[width=\linewidth]{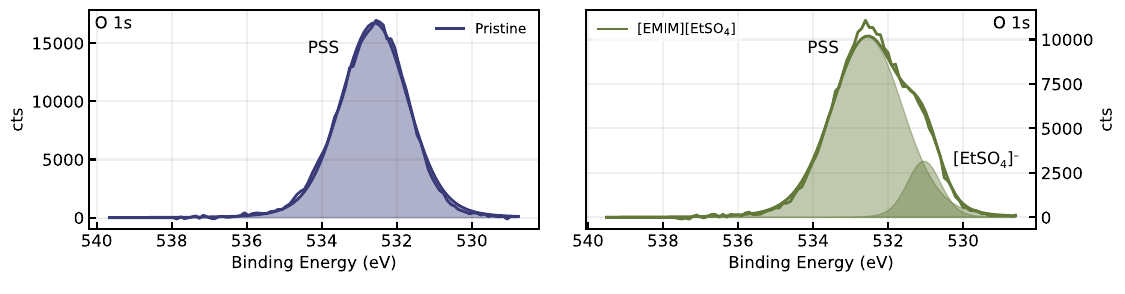} 
    \caption{\textbf{XPS spectra (O\,1s) of PEDOT:PSS for ion exchange estimation.} XPS (O\,1s) spectra and fits of pristine PEDOT:PSS and after exposure to \ce{[EMIM][EtSO4]}. \label{fig:S_XPS_O}}
     \vspace*{-\baselineskip}
\end{figure}

\begin{equation}
    R_\mathrm{IE} = R_\mathrm{M} \cdot \frac{\int{\mathrm{O\,1s}(\ce{[EtSO4]-})\,\mathrm{d}E}}{\int{\mathrm{O\,1s}(\ce{PSS})\,\mathrm{d}E}} = R_\mathrm{M} \cdot \frac{\mathrm{cts}^{\mathrm{O\,1s}}(\ce{[EtSO4]-})}{\mathrm{cts}^{\mathrm{O\,1s}}(\ce{PSS})}
\end{equation}

\begin{table}[h!]
\caption{\textbf{Ion-exchange ratios estimated from XPS.}}
\centering
\begin{tabular}{r l l l l}
\hline
System &  $\mathrm{cts}^{\mathrm{O\,1s}}(\ce{PSS})$ & $\mathrm{cts}^{\mathrm{O\,1s}}(\ce{[EtSO4]-})$ & $R_\mathrm{M}$ & $R_\mathrm{IE}$\\
\hline
Pristine & $38907.29$ &   &  $3.39$ \\
\ce{[EMIM][EtSO4]} & $25046.53$ &   $3694.77$ & $1.61$ & $0.24$\\
\hline
\end{tabular}
\label{tab:O_XPS_Ratios}
\end{table}
\noindent We calculate the ion-exchange ratio as approximately 24\%. This figure does not deviate significantly from the estimate in \hyperref[Note_S:Ion_Transfer]{Supplementary Note 2}, where a theoretical exchange ratio of 31.7\% is derived from DFT literature data.
\\ \\ Since we infer the ion-exchange ratio by relating information from the S\,2p core level region to the O\,1s region, we verify this approach by comparing the ratio of \ce{PEDOT}-equivalents for the pristine and \ce{[EMIM][EtSO4]}-treated sample in the O\,1s and S\,2p spectra:
\begin{align}
    \frac{\mathrm{cts}^{\text{O\,1s}}_{\text{Pristine}}(\ce{PSS})}{\mathrm{cts}^{\text{O\,1s}}_{\ce{[EMIM][EtSO4]}}(\ce{PSS})} \cdot \frac{R_\mathrm{M}^{\ce{[EMIM][EtSO4]}}}{R_\mathrm{M}^{\text{Pristine}}} & = 0.74 \\
    \frac{\mathrm{cts}^{\text{S\,2p}}_{\text{Pristine}}(\ce{PEDOT})}{\mathrm{cts}^{\text{S\,2p}}_{\ce{[EMIM][EtSO4]}}(\ce{PEDOT})} &= 0.70
\end{align}
The deviation between the two ratios is only around 6\%, which we consider small enough to motivate our previous calculations. 
\\ \\ 
\noindent Fig.\,\ref{fig:S_Composition} illustrates the channel compositions of pristine PEDOT:PSS and channels exposed to \ce{[EMIM][TFSI]} and \ce{[EMIM][EtSO4]}.

\section*{Supplementary Note 4: DFT Calculations}
\addcontentsline{toc}{subsection}{Supplementary Note 4: DFT Calculations}
\label{Note_S:Raman_DFT}

To understand the impact of intra-chain structure on the measured Raman spectra, PEDOT oligomers consisting of 10 repeat units and three holes were studied using density functional theory (DFT). Using oligomers of ten repeat units allowed for appropriate delocalization of excess charges and charge densities for doped PEDOT\autocite{kahol2005metallic}, while also allowing for structures of representative sizes\autocite{kim2021long, jain2021pedot} to be simulated. All simulations were performed using doublet multiplicity. 

Simulations were performed using two distinct initial configurations of PEDOT. The first with all inter-thiophene dihedrals set to zero, and the second with the central inter-thiophene dihedral set to $140\si{\degree}$. These two structures were geometry relaxed using DFT with a B3LYP\autocite{becke1988density, lee1988development, slater1974self} functional and 6-311G basis set.  Following the geometry relaxation, a frequency analysis was performed to obtain their vibrational frequencies corresponding Raman activities, allowing for the dependency on the frequency of the incident Raman laser. An incident laser energy of $2.329\,\si{\electronvolt}$ was used, matching that of the experiments. DFT calculations were carried out using the Gaussian16 software package\autocite{g16}.  

The Raman spectra were obtained by fitting Lorentzian functions to the Raman frequencies, each with a maximum corresponding to the Raman activity and a full-width-half-maximum of $5\,\si{\per\centi\metre}$. 

\section*{Supplementary Note 5: Absorbance Fits for Spectroelectrochemistry}
\addcontentsline{toc}{subsection}{Supplementary Note 5: Absorbance Fits for Spectroelectrochemistry}
\label{Note_S:Spectroelectrochemistry_Fits}

\begin{figure}[b]
    \centering
    \includegraphics[width=\linewidth]{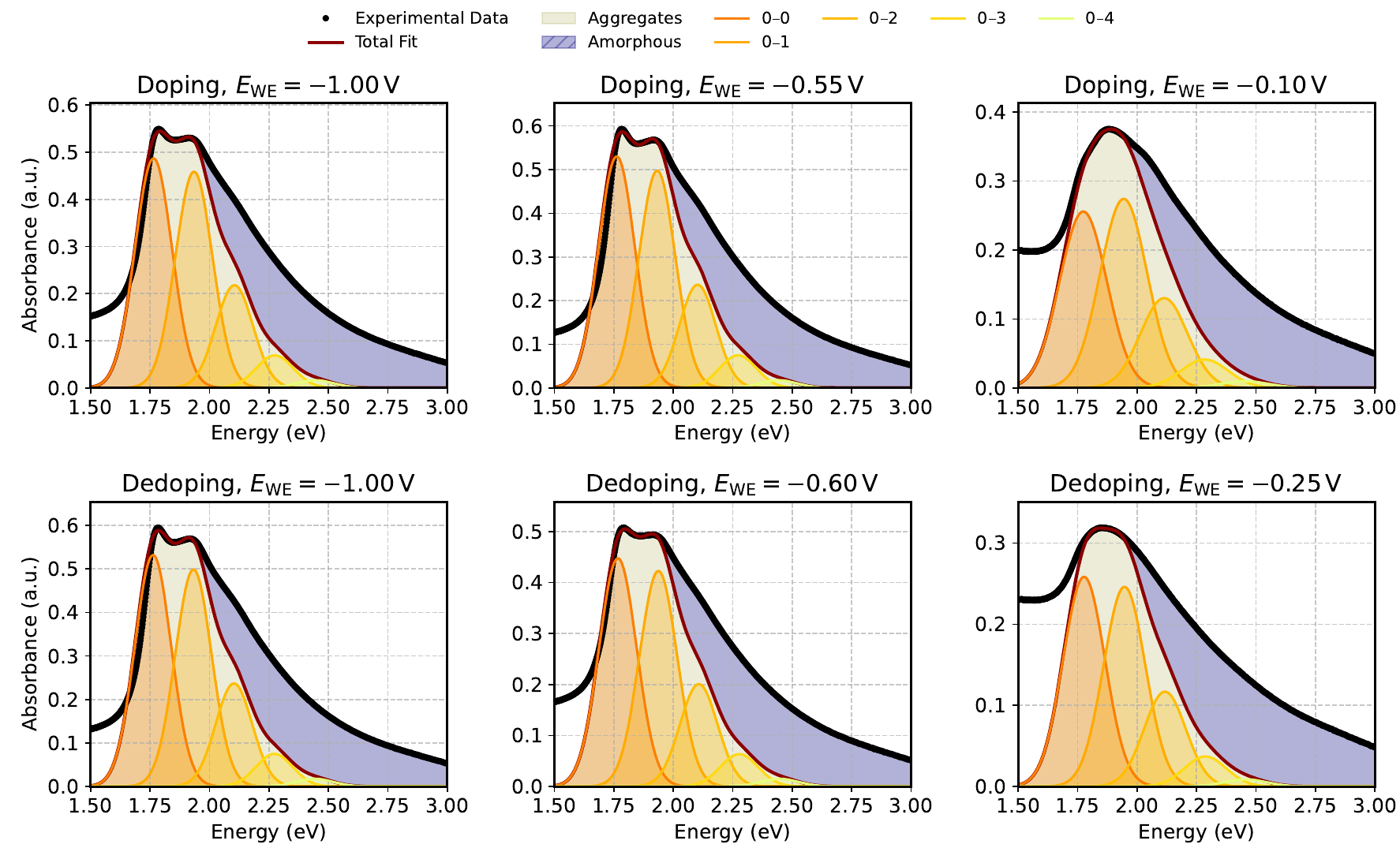} 
    \caption{\textbf{Representative vibronic fits.} Fits of the Holstein-Spano model for the sample with \ce{[EMIM][EtSO4]} electrolyte.
    \label{fig:S_fit_results}}
     \vspace*{-\baselineskip}
\end{figure}

Absorbance (Abs.) in this work refers to
\begin{equation}
    \mathrm{Abs.} = -\log{\left(\frac{I}{I_0}\right)},
\label{eq:S_absorbance}
\end{equation}
where $I$ and $I_0$ are the intensities of light transmitted through sample and reference. Eq.\,\ref{eq:S_absorbance} directly relates to the oscillator strength of a transition by the Beer-Lambert law. We fit the optical absorbance spectra to the vibronic model developed by Spano\autocite{spano2005modeling}, using methods previously shown applicable for other thiophene-based polymers\autocite{lecroy2023role}. The model takes a Holstein Hamiltonian as a basis and assumes a molecular crystal with electron-phonon coupling by a single phonon mode. Absorbance results as a function of energy (frequency $\omega$) by
\begin{equation}
\label{eq:S_Spano_Absorbance}
       A(\omega) = C \sum_{m=0} \frac{e^{-\lambda^2} \lambda^{2m}}{m!} 
\left(1 - \frac{W e^{-\lambda^2}}{2 E_p} G(\lambda^2; m) \right)^2 
\Gamma\left(\hbar\omega - E_{00} - m E_p - 0.5 W \lambda^{2m} e^{-\lambda^2}\right)
\end{equation}
with
\begin{align}
        G(\lambda^2; m) &= \sum_{\substack{n=0,1\ldots \\ n\neq m}}\frac{\lambda^{2n}}{n! (n - m)} \quad \text{and} \\
        \Gamma(X) &= \frac{1}{\sigma\sqrt{2\pi}} e^{-\frac{X^2}{2\sigma^2}},
\end{align}
where $\hbar$ is the reduced Planck constant. Fits are performed with four free parameters, being the scaling factor $C$, the exciton bandwidth $W$, the 0--0 vibronic transition energy $E_{00}$, and the peak linewidth $\sigma$. For $W$, $W = 4 J_\mathrm{inter}$ applies, with $J_\mathrm{inter}$ denoting interchain exciton coupling. $\lambda^2$ is the Huang-Rhys factor, essentially denoting the strength of electron-phonon coupling. It is taken here as $\lambda^2=0.95$, based on previous works on thiophene-based polymers\autocite{lecroy2023role, clark2007role}. $E_p$ is a vibrational energy combining multiple phonon modes stemming from the thiophene backbone and was kept at $E_p=0.17\,\si{\volt}$, based on the $1367\,\si{\per\centi\metre}$ Raman mode (Tab.\,\si{\ref{tab:S_Raman}}). We consider the energetically five lowest vibronic transitions (0--0 to 0--4), thus taking the absorbance sum from $m=0$ to $m=4$. Fits were performed by a Levenberg-Marquardt algorithm with the modifications introduced by Fletcher\autocite{fletcher1971modified}. Fig.\,\ref{fig:S_fit_results} shows representative fit results for the sample with \ce{[EMIM][EtSO4]} electrolyte.

\begin{figure}[H]
    \centering
    \includegraphics[width=0.6\linewidth]{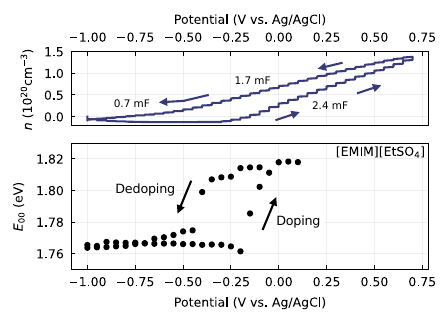} 
    \caption{\textbf{Ground state transition energies from manual fits.} When adjusting the fitting range manually, the energy shift in $E_{00}$ amounts to $\sim53\,\si{\milli\electronvolt}$.
    \label{fig:E_00_note}}
     \vspace*{-\baselineskip}
\end{figure}

\noindent The aggregate fraction ($AF$) is calculated from the integrated spectral intensities as estimated from fitting Eq.\,\ref{eq:S_Spano_Absorbance}. These are assigned to aggregate absorbance $A_\mathrm{Agg}$, while the remainder to the total experimental absorbance is assigned to amorphous regions $A_\mathrm{Amor}$. $AF$ results as the ratio of integrated aggregate absorbance to the integrated absorbance of both aggregates and amorphous regions, weighted by the molar extinction coefficients $\epsilon_\mathrm{Agg}$ and $\epsilon_\mathrm{Amor}$:
\begin{equation}
    AF = \frac{\epsilon_\mathrm{Agg}\int A_\mathrm{Agg}(E)\,\mathrm{d}E}{\epsilon_\mathrm{Agg}\int A_\mathrm{Agg}(E)\,\mathrm{d}E + \epsilon_\mathrm{Amor}\int A_\mathrm{Amor}(E)\,\mathrm{d}E}
\end{equation}
Integrals are taken in a range of $\sim1.13$--$3.10\,\si{\electronvolt}$ $(400$--$1100\,\si{\nano\metre})$, corresponding to our setup's detector limits. $\epsilon_\mathrm{Agg}$ and $\epsilon_\mathrm{Amor}$ are adopted from values reported on poly(3-hexylthiophen-2,5-diyl) (P3HT) with $\epsilon_\mathrm{Agg}=1.39\epsilon_\mathrm{Amor}$\autocite{clark2009determining}. 
\\ \\
\noindent Worth noting, we find that when evaluating the spectrum visually and adjusting the fitting range manually to best match the experimental data, we find an even larger shift in $E_{00}$, namely by about $53\,\si{\milli\electronvolt}$ (Fig.\,\ref{fig:E_00_note}). However, for reproducibility reasons, we here focus on the ones obtained from an automated least-square algorithm. 

\clearpage
\section*{Supplementary References}
\printbibliography[heading = none]